\documentclass[11pt,a4paper]{article}
\pdfoutput=1

\usepackage[utf8]{inputenc}
\usepackage{a4wide}
\usepackage{lmodern}
\usepackage{amsmath}
\usepackage{cite}
\usepackage{xcolor}
\usepackage{amssymb}
\usepackage{amsfonts}
\usepackage{booktabs}
\usepackage{makecell}
\usepackage{multirow}
\usepackage{pdflscape}
\usepackage{textcomp}
\usepackage{gensymb}
\usepackage{bigstrut}
\usepackage{array}
\usepackage{enumerate}
\usepackage[small,bf]{caption}
\usepackage{graphicx}
\usepackage{hyperref}
\usepackage{mathbbol}
\usepackage{amssymb}
\usepackage{appendix}
\usepackage{verbatim}
\usepackage{geometry}
\usepackage{makecell}
\usepackage{listings}
\usepackage{mathtools}
\usepackage{dsfont}
\usepackage{accents}
\usepackage{longtable}
\usepackage{cleveref}

\DeclareMathOperator{\diag}{diag}

\newcommand{\id}{\mathds{1}}

\newcommand\barparenn{\overset{\textbf{\fontsize{4pt}{4pt}\selectfont(---)}}{\nu}}
\newcommand\barparene{\overset{\textbf{\fontsize{4pt}{4pt}\selectfont(---)}}{\nu}_{\!\!e}}
\newcommand\barparenm{\overset{\textbf{\fontsize{4pt}{4pt}\selectfont(---)}}{\nu}_{\!\!\mu}}
\newcommand\barparena{\overset{\textbf{\fontsize{4pt}{4pt}\selectfont(---)}}{\nu}_{\!\!\alpha}}

\newcommand*{\belowrulesepcolor}[1]{%
  \noalign{%
    \kern-\belowrulesep
    \begingroup
      \color{#1}%
      \hrule height\belowrulesep
    \endgroup
  }%
}
\newcommand*{\aboverulesepcolor}[1]{%
  \noalign{%
    \begingroup
      \color{#1}%
      \hrule height\aboverulesep
    \endgroup
    \kern-\aboverulesep
  }%
}
\definecolor{light-gray}{gray}{0.9}
\definecolor{lighter-gray}{gray}{0.95}

\allowdisplaybreaks
\numberwithin{equation}{section}

\begin{document}
\begin{titlepage}

\vspace*{-15mm}
\begin{flushright}
CFTP/22-003
\end{flushright}
\vspace*{5mm}

\begin{center}
{\bf\LARGE {
Baseline and other effects\\[3mm] for a sterile neutrino at DUNE}
}\\[8mm]

J.~T.~Penedo\(^{\,}\)\footnote{E-mail: \texttt{joao.t.n.penedo@tecnico.ulisboa.pt}} and
João~Pulido\(^{\,}\)\footnote{E-mail: \texttt{pulido@cftp.ist.utl.pt}}\\
 \vspace{5mm}
{\it CFTP, Departamento de Física, Instituto Superior Técnico, Universidade de Lisboa,\\
Avenida Rovisco Pais 1, 1049-001 Lisboa, Portugal}
\end{center}
\vspace{2mm}

\begin{abstract}
We analyze the sensitivity of the Deep Underground Neutrino Experiment \linebreak (DUNE) to a sterile neutrino, combining information from both near and far detectors. We quantify often-neglected effects which may impact the event rate estimation in a 3+1 oscillation scenario. In particular, we find that taking into account the information on the neutrino production point, in contrast to assuming a pointlike neutrino source, affects DUNE's sterile exclusion reach. Visible differences remain after the inclusion of energy bin-to-bin uncorrelated systematics. Instead, implementing exact oscillation formulae for near detector events, including a two slab density profile, does not result in any visible change in the sensitivity.
\end{abstract}

\end{titlepage}
\setcounter{footnote}{0}

\section{Introduction}
\label{sec:intro}

Sterile neutrinos are hypothetical neutral leptons that interact with other particles via gravity, missing all the gauge interactions of the Standard Model. They can only be indirectly detected through their mixing with standard weakly-interacting neutrinos. 

The possible existence of light sterile neutrinos with a mass of order 1 eV (for recent reviews see~\cite{Giunti:2019aiy,Boser:2019rta,Dasgupta:2021ies}) has been strongly hinted at in the mid-1990s with the results of the LSND accelerator experiment~\cite{Athanassopoulos:1995iw,Athanassopoulos:1996jb}, which found evidence in favor of short-baseline (SBL) $\bar\nu_{\mu}\rightarrow \bar\nu_{e}$ oscillations ($\bar\nu_{e}$ appearance).
The $\bar\nu_e$ excess in the data indicated an oscillation to a sterile neutrino with a mass-squared difference to the interacting ones of $\Delta m^{2}_{41}\gtrsim 0.1 \text{ eV}^2$. The MiniBooNE experiment~\cite{AguilarArevalo:2007it}, developed later with the main purpose of testing the LSND signal and initially operating in neutrino mode, reached results which could not confirm the effect but, on the other hand, appeared to hint at additional effects in the low-energy part of the spectrum.
More recently, new data from MiniBooNE operating in neutrino and antineutrino modes are consistent in energy and magnitude with the excess of events reported by LSND and lead to a combined LSND/MiniBooNE best fit at $\Delta m^2_{41}=0.041\text{ eV}^2$ and $\sin^2 2\theta_{e\mu}=\sin^2 2\theta_{14}\,\sin^2 \theta_{24}=0.96$~\cite{Aguilar-Arevalo:2018gpe}.
This result is, however, not confirmed by the OPERA~\cite{Agafonova:2018dkb} accelerator experiment. Moreover, large values of the mixing angles between active and sterile are in strong tension with solar neutrino data~\cite{Giunti:2009xz,Palazzo:2011rj,Palazzo:2012yf}, as well as with the results from disappearance experiments, as we shall soon discuss.%
\footnote{
Cosmological data provides a very stringent independent constraint $|U_{\alpha 4}|^2 \lesssim 10^{-3}$ ($\alpha=e,\mu,\tau$)~\cite{Hagstotz:2020ukm}. The discovery of sterile-neutrino mixing at the levels indicated by SBL hints would thus need to be reconciled with cosmology (see e.g.~\cite{Hannestad:2013ana,Chu:2018gxk}).}
Owing to the unclear situation, a new set of experiments --- the Short Baseline Neutrino program at Fermilab --- was proposed some time ago and is being developed to check the LSND and MiniBooNE $\bar\nu_{e}$ excesses~\cite{Antonello:2015lea,Tufanli:2017mwt}.

Since the release of the initial MiniBooNE results, the light sterile neutrino issue remained dormant, until it was revived in 2011 with the discovery of the reactor antineutrino anomaly~\cite{Mueller:2011nm,Mention:2011rk,Huber:2011wv}. 
In particular, reassessed $\bar\nu_e$ fluxes were found to imply larger detection rates with respect to what was measured in several SBL reactor neutrino experiments ($\bar\nu_e$ disappearance), indicating a deficit in the data. This deficit could be made consistent with oscillations generated by $\Delta m^2_{41}\gtrsim 0.5\text{ eV}^2$~\cite{Giunti:2019aiy}.
Further attention was also given to the previously-known Gallium neutrino anomaly~\cite{Giunti:2006bj,Laveder:2007zz,Giunti:2010zu,Giunti:2012tn}. In this context, a deficit of neutrinos originating from the intense $^{51}$Cr and $^{37}$Ar neutrino sources was found in the GALLEX~\cite{Kaether:2010ag} and SAGE~\cite{Abdurashitov:2005tb} detectors ($\nu_e$ disappearance) during calibration. The Gallium anomaly can be explained through oscillations generated by a sterile with $\Delta m^2_{41}\gtrsim 1\text{ eV}^2$~\cite{Giunti:2019aiy}.
While recent updates to the flux calculation may have possibly resolved the reactor anomaly~\cite{Estienne:2019ujo,Kopeikin:2021ugh,Giunti:2021kab}, the case for a Gallium anomaly has been strengthened by the latest results from BEST~\cite{Barinov:2021asz}. In a 3+1 interpretation, the combination of all Gallium data yields the best-fit point $\Delta m^2_{41}=1.25\text{ eV}^2$ and $\sin^2 \theta_{14}=9.4\times 10^{-2}$ (see also~\cite{Barinov:2021mjj}).

In order to investigate whether neutrino oscillations are the source of the aforementioned anomalies, one may measure the antineutrino flux from the reactor core at different distances, thus obtaining ratios of event rates. In this way, flux normalization and energy-shape uncertainties could be reduced. Along these lines, interesting results have been obtained by the NEOS experiment~\cite{Ko:2016owz}. Their detector is, however, located at a fixed distance from the reactor and their data were compared with those from a similar reactor at Daya Bay at a much larger distance. This was followed by the DANSS experiment~\cite{Alekseev:2018efk}, measuring the flux from one single reactor core at different locations. Initially, there appeared to be a remarkable overlap of allowed regions of the two sets of data, indicating~\cite{Dentler:2018sju}
\begin{equation}
  \Delta m^2_{41}\simeq 1.3\text{ eV}^2\,, \qquad\sin^2 \theta_{14} \simeq 0.01\,.
  \label{eq:fit1}
\end{equation}
More recent results from DANSS have, however, exhibited a smaller significance~\cite{Danilov:2019aef,Danilov:2020ucs} (see also~\cite{Licciardi:2021hyi}).
The point of \cref{eq:fit1} is not only in disagreement with the LSND and MiniBooNE preferred regions but also with the BEST joint Gallium fit.
Other experiments~\cite{Almazan:2018wln,Ashenfelter:2018iov,Serebrov:2018vdw,Abreu:2018njy} are also dedicated to measuring the $\bar\nu_{e}$ reactor flux as a function of a varying (short) baseline. These, however, expect much smaller inverse-beta-decay event rates and signal-to-background ratios with respect to NEOS and DANSS (see, e.g.,~\cite{Danilov:2018dme,Danilov:2022str}). Of these experiments, Neutrino-4 has found an unexpected indication of oscillations corresponding to $\Delta m^2_{41}\simeq 7\text{ eV}^2$ and $\sin^2\theta_{14} \simeq 0.1$~\cite{Serebrov:2018vdw}, which does not overlap with the regions preferred by reactor data. The Neutrino-4 result has been reexamined~\cite{Giunti:2021iti}, and its statistical significance remains a source of controversy. 

If light sterile neutrinos exist and mix with the active ones, then they are expected to provide a signature in $\nu_{\mu}$ and $\bar\nu_{\mu}$ disappearance. So far, no such effect has been observed. A stringent upper bound on the corresponding 3+1 mixing angle $\theta_{24}$ was obtained by combining 
MINOS and MINOS+ data~\cite{Adamson:2017uda},
\begin{equation}
  \sin^2\theta_{24}\lesssim 10^{-2}
  \,\,(90\%\text{ C.L.})\,,
  \label{eq:bound1}
\end{equation}
for $\Delta m^2_{41}\gtrsim 10^{-2}\text{ eV}^2$. The preferred region for the sterile neutrino parameters is still unclear. 
There is, in fact, a strong appearance-disappearance tension that can be quantified through the violation of the approximate relation
\begin{equation}
  \sin^2 2\theta_{e\mu}\simeq \frac{1}{4} \sin^2 2\theta_{ee}~\sin^2 2\theta_{\mu\mu}\,,
  \label{eq:approx}
\end{equation}
where $\sin^2 2\theta_{e\mu}$ governs $\barparene$ appearance, while $\sin^2 2\theta_{ee} = \sin^2 2\theta_{14}$ and $\sin^2 2\theta_{\mu\mu} \simeq \sin^2 2\theta_{24}$ (for small angles) control $\barparene$ and $\barparenm$ disappearance, respectively.
The disagreement between the data and \cref{eq:approx} becomes clear if one inserts in the left-hand side the combined lower bound from SBL $\barparenm\rightarrow \barparene$ experiments which include LSND~\cite{Athanassopoulos:1995iw}, KARMEN~\cite{Armbruster:2002mp}, OPERA~\cite{Agafonova:2018dkb}, NOMAD~\cite{Astier:2003gs}, BNL-E776~\cite{Borodovsky:1992pn} and ICARUS~\cite{Antonello:2013gut}, namely~\cite{Giunti:2019aiy}
\begin{equation}
  \sin^2 2\theta_{e\mu}\gtrsim 10^{-3}
   \,\,(3\sigma\text{ C.L.})\,,
\end{equation}
and in its right-hand side the indication and bound of \cref{eq:fit1,eq:bound1}, resulting in
\begin{equation}
 \frac{1}{4} \sin^2 2\theta_{ee}~\sin^2 2\theta_{\mu\mu}\lesssim 4\times 10^{-4}\,.
\end{equation}
%
Although these are approximate values, they are sufficient to illustrate the existing tension between appearance and disappearance data. 

Altogether, the question of the possible existence of sterile neutrinos and their intrinsic parameters is therefore far from settled. To this end, the upcoming Deep Underground Neutrino Experiment (DUNE)~\cite{Abi:2020wmh,Abi:2020evt,Abi:2020oxb,Abi:2020loh} may play an important clarifying role.
In the DUNE setup, whose nominal mission is to perform precise measurements of neutrino properties and oscillations, a neutrino beam will be obtained from meson decays within a 194 m long decay pipe, which is followed by a hadron absorber. Originating from a proton beam impinging on a graphite target, positive and negative mesons are to be focused by a three-horn system~\cite{DUNE:2021cuw}, which can be operated either in forward (FHC) or reverse (RHC) horn current mode, leading to the production of a beam of mostly neutrinos or antineutrinos, respectively.
Downstream from the graphite target, at a distance of approximately 574 m, a liquid-argon detector is to be installed as part of the near detector (ND) complex~\cite{DUNE:2021tad}. Further, a second liquid-argon far detector (FD) with a fiducial mass of around 40 kt is to be located at the Sanford Underground Research Facility in South Dakota, at a distance of 1297 km from Fermilab.

The purpose of this paper is to analyze the prospects of DUNE in constraining the sterile-neutrino parameter space, quantifying effects which may impact event rate estimations.
We focus on the range $\Delta m^2_{41} \in [0.1,100]$ eV$^2$ for the sterile mass-squared difference. For this range of the new mass-squared difference and depending on the (anti)neutrino energy, the associated oscillation length may be of the order of the decay pipe length and of the distance between the graphite target and the ND complex. Thus, the location of the neutrino production point within the decay pipe can play an important role in determining event rates at the ND, as it influences the oscillation probabilities by affecting the oscillation baseline.
This smeared-source effect goes beyond taking into account geometrical effects in the computation of unoscillated fluxes at the ND (present also in the 3+0 case) and has not been quantitatively assessed in previous works.%
\footnote{In ref.~\cite{Abi:2020kei} an additional 20\% Gaussian energy smearing has been considered in its place.}
Note that the prospects for detecting light sterile neutrinos in the DUNE ND have been examined in refs.~\cite{Choubey:2016fpi,Blennow:2016jkn,Tang:2017khg,Miranda:2018yym,Abi:2020kei,Coloma:2021uhq,Ghosh:2021rtn}.%
\footnote{The effects in the DUNE ND of heavier sterile neutrino states, in the MeV-GeV mass range (so-called heavy neutral leptons), have been analyzed in refs.~\cite{Krasnov:2019kdc,Ballett:2019bgd,Berryman:2019dme,Breitbach:2021gvv,Carbajal:2022zlp}.
}
Our analysis improves on the existing literature as it takes into account ND and FD event rates, shape uncertainties, exact 3+1 oscillation formulae in matter and, crucially, the aforementioned baseline dependence, while making use of the latest DUNE configurations~\cite{DUNE:2021cuw}.

The paper is organized as follows. In \cref{sec:framework} we present the theoretical framework to be used. In \cref{sec:dune} we expound the details of the DUNE simulation and present our main results. These are represented by discrepancies between exclusion curves in the sterile parameter space, which primarily arise as one takes into account the effect of a spatially-smeared source. It is found that such discrepancies persist after taking into account energy bin-to-bin uncorrelated systematic uncertainties. In \cref{sec:summary} we summarize our conclusions.

\section{Neutrino oscillations in vacuum and matter}
\label{sec:framework}

We work within the 3+1 framework, in which the $3\nu$-flavor paradigm is extended by a single sterile neutrino species $\nu_s$. In this framework, 
the propagation of a neutrino in the Earth's matter is described by the Hamiltonian ($\alpha, \beta = e,\mu,\tau,s$)
\begin{equation}
\begin{split}
    H_{\alpha\beta}^\text{mat} &= \frac{1}{2E} \left[U \begin{pmatrix}
    0 &0 &0 &0 \\
    0 &\Delta m_{21}^2 &0 &0 \\
    0 &0 &\Delta m_{31}^2 &0 \\
    0 &0 &0 &\Delta m_{41}^2
    \end{pmatrix}
    U^\dagger 
    +
    \begin{pmatrix}
    A_\text{CC} &0 &0 &0 \\
    0 &0 &0 &0 \\
    0 &0 &0 &0 \\
    0 &0 &0 &A_\text{NC}
    \end{pmatrix}\right]_{\alpha\beta}
    \!\!\!,
\end{split}
\label{eq:H}
\end{equation}
where $U$ is the vacuum Pontecorvo-Maki-Nakagawa-Sakata (PMNS) lepton mixing matrix, $\Delta m_{ij}^2 \equiv m_i^2 - m_j^2$ are vacuum mass-squared differences, and $E$ denotes the energy of the neutrino. The coherent forward elastic charged current (CC) scattering contribution $A_\text{CC} = 2\sqrt{2}\,G_F N_e E$ depends on the density $N_e$ of electrons in the medium.
The net effect of neutral current (NC) scatterings is encoded in $A_\text{NC} =  \sqrt{2}\,G_F N_n E$. In the Earth, only the neutron density $N_n \simeq N_e$ is relevant as the NC potentials of protons and electrons cancel. One has
\begin{equation}
    A_\text{CC} \,\simeq\, 7.6 \times 10^{-5}  \left(\frac{\rho}{\text{g\,cm}^{-3}}\right) \left(\frac{E}{\text{GeV}}\right) \,\text{eV}^2 \,,
    \quad
    A_\text{NC} \,\simeq\, \frac{1}{2} A_\text{CC}\,.
\end{equation}
To study the propagation of antineutrinos in matter, it is sufficient to replace $U \to U^*$, $A_\text{CC} \to -A_\text{CC}$ and $A_\text{NC}\to -A_\text{NC}$ in \cref{eq:H}, see, e.g.,~\cite{Akhmedov:2004ny}.
The vacuum Hamiltonian is recovered when $\rho = 0$,
\begin{equation}
\begin{split}
    H_{\alpha\beta}^\text{vac} &= \frac{1}{2E} \left[U \begin{pmatrix}
    0 &0 &0 &0 \\
    0 &\Delta m_{21}^2 &0 &0 \\
    0 &0 &\Delta m_{31}^2 &0 \\
    0 &0 &0 &\Delta m_{41}^2
    \end{pmatrix}
    U^\dagger 
    \right]_{\alpha\beta}
    \!\!\!.
\end{split}
\label{eq:Hvac}
\end{equation}
The eigenvalues of $H^\text{vac}$ are proportional to the vacuum mass-squared differences and read $\Delta m^2_{i1} / 2E$. 

The matter Hamiltonian of \cref{eq:H} can be brought to the form of \cref{eq:Hvac}, 
\begin{equation}
\begin{split}
    H_{\alpha\beta}^\text{mat} &= 
    \frac{1}{2E} \left[ \tilde{U} \begin{pmatrix}
    0 &0 &0 &0 \\
    0 &\Delta \tilde{m}_{21}^2 &0 &0 \\
    0 &0 &\Delta \tilde{m}_{31}^2 &0 \\
    0 &0 &0 &\Delta \tilde{m}_{41}^2
    \end{pmatrix}
    \tilde{U}^\dagger
    + \hat\Delta m^2_{11}\id
    \right]_{\alpha\beta}
    \!\!\!,
\end{split}
\label{eq:H2}
\end{equation}
up to the constant term $\hat \Delta m_{11}^2/2E$, which, while non-zero in general, does not impact oscillation probabilities. We denote the eigenvalues of the matter Hamiltonian by $\hat \Delta m^2_{i1} / 2E$ ($i=1,...,4$), with $\hat \Delta m_{i1}^2 = \Delta \tilde{m}_{i1}^2 + \hat \Delta m_{11}^2$ (see also \mbox{\cref{app:diag}}).
The oscillation parameters in matter, denoted with tildes, are defined by \cref{eq:H2}.
Neutrino oscillations in matter are sensitive to the matter mass-squared differences $\Delta \tilde m_{ij}^2 \equiv \Delta \tilde{m}_{i1}^2 - \Delta \tilde{m}_{j1}^2$. 
Both vacuum and matter PMNS mixing matrices $U$ and $\tilde{U}$ can be parametrized in terms of six mixing angles and three CP violation (CPV) phases,%
\footnote{We disregard Majorana phases since they do not play a role in neutrino oscillations in vacuum~\cite{Bilenky:1980cx,Doi:1980yb} or in matter~\cite{Langacker:1986jv}, as can be inferred from \cref{eq:H}.}
\begin{align}
    U &\,=\,
    R_{34}(\theta_{34},\delta_{34})\,
    R_{24}(\theta_{24},\delta_{24})\,
    R_{14}(\theta_{14})\,
    R_{23}(\theta_{23})\,
    R_{13}(\theta_{13},\delta_{13})\,
    R_{12}(\theta_{12})\,,
\label{eq:paramU}\\[1mm]
    \tilde{U} &\,=\,
    R_{34}(\tilde\theta_{34},\tilde\delta_{34})\,
    R_{24}(\tilde\theta_{24},\tilde\delta_{24})\,
    R_{14}(\tilde\theta_{14})\,
    R_{23}(\tilde\theta_{23})\,
    R_{13}(\tilde\theta_{13},\tilde\delta_{13})\,
    R_{12}(\tilde\theta_{12})\,,
\label{eq:paramUtilde}
\end{align}
with $R_{ij}(\theta) = R_{ij}(\theta,0)$.
The nontrivial $(i,j)$ block of each complex rotation $R_{ij}$ follows the convention
\begin{equation}
    R_{ij}(\theta,\delta)\big|_{(i,j)} = \begin{pmatrix}
    \cos\theta & \sin\theta\, e^{-i\delta} \\
    -\sin\theta\, e^{i\delta} & \cos\theta
    \end{pmatrix}
\end{equation}
and \cref{eq:paramU,eq:paramUtilde} reduce to the standard parametrization~\cite{Zyla:2020zbs} in the limits of vanishing sterile mixing angles, $\theta_{i4} = 0$ and $\tilde\theta_{i4} = 0$ ($i=1,2,3$).

The probability $P_{\alpha\beta}$ of transition between neutrino flavors $\alpha$ and $\beta$ (or of survival for a given flavor, $\alpha = \beta$) after traversing a length $L$ under the influence of a Hamiltonian $H$ is given by
\begin{equation}
    P_{\alpha\beta}=|\langle \nu_\beta | \nu_\alpha(L)\rangle|^2
    = |\langle \nu_\beta| e^{-i H L}| \nu_\alpha\rangle|^2
    \,.
\label{eq:P}
\end{equation}
For propagation in vacuum, the eigenstates $|\nu_k\rangle$ of $H=H^\text{vac}$ are related to the flavor states via $|\nu_\alpha\rangle = U_{\alpha k}^* |\nu_k\rangle$. By inserting the completeness relation $\sum_k |\nu_k\rangle\langle\nu_k|= 1$ in \cref{eq:P}, one obtains the known formula for the vacuum oscillation probabilities, 
\begin{equation}
\begin{aligned}
P_{\alpha\beta}^\text{vac}(L,E)
\,=\,
\delta_{\alpha\beta}
&-   4 \sum_{i>j}\,\text{Re}
\left[U_{\alpha i}^*\,U_{\beta i}\,U_{\alpha j}\,U_{\beta j}^*\right]
\,\sin^2 \Delta_{ij} \\
&\pm 2 \sum_{i>j}\,\text{Im}
\left[U_{\alpha i}^*\,U_{\beta i}\,U_{\alpha j}\,U_{\beta j}^*\right]
\,\sin 2 \Delta_{ij} \,,
\end{aligned}
\label{eq:Pvac}
\end{equation}
where $i,j = 1,2,3,4$, and the upper (lower) sign choice refers to (anti)neutrinos. This follows since, for antineutrinos, $|\bar\nu_\alpha\rangle = U_{\alpha k} |\bar\nu_k\rangle$. One has here defined
\begin{align}
\Delta_{ij} \,\equiv \, \frac{\Delta m^2_{ij}\, L}{4E}
\,\simeq\, 1.27\,\frac{\Delta m_{ij}^{2}\,[\text{eV}^{2}]\,L\,[\text{km}] }{ E\,[\text{GeV}]}\,.
\end{align}
Instead, for propagation in a medium of constant density ($\rho \neq 0$), the eigenstates $|\tilde \nu_k\rangle$ of $H=H^\text{mat}$ obey $|\nu_\alpha\rangle = \tilde U_{\alpha k}^* |\tilde\nu_k\rangle$. The probabilities $P_{\alpha\beta}^\text{mat}$ are then given by \mbox{\cref{eq:Pvac}} with the straightforward replacements $U \to \tilde U$ and $\Delta_{ij} \to \tilde \Delta_{ij} \equiv \Delta \tilde m^2_{ij} L/4E$.

\vskip 2mm

\begin{figure}[t]
\centering
\includegraphics[width=\textwidth]{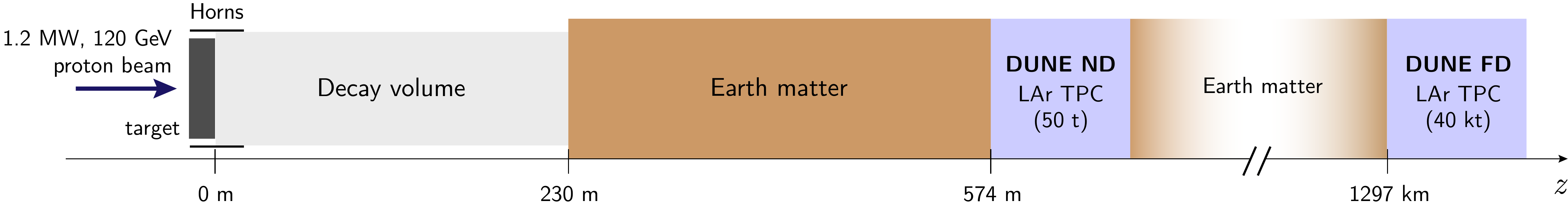}
\caption{Diagram of the DUNE beam setup, where $z$ represents the distance from the graphite target (adapted from~\cite{Berryman:2019dme}).}
\label{fig:setup}
\end{figure}

We are interested in the DUNE scenario, with an emphasis on ND events. The relevant setup is shown in \Cref{fig:setup}.
One sees that the distance traveled by the neutrinos from the production point to the ND consists of a first segment of (event-dependent) length $L_1 < 230$ m in a low-density medium ($\rho\simeq 0$), which includes the 194 m decay pipe, and a second segment of fixed length $L_2\simeq 344$ m in the Earth's crust with an approximately constant density, $\rho\simeq 2.6$ g cm$^{-3}$.
In a two slab description of the matter profile, using the Hamiltonians in \cref{eq:Hvac,eq:H2}, the short-baseline oscillation probabilities read
\begin{equation}
\begin{split}
    P_{\alpha\beta}^\text{SBL} &= \left|\,\langle \nu_\beta| e^{-i H^\text{mat} L_2}\, e^{-i H^\text{vac} L_1}| \nu_\alpha\rangle \,\right|^2
    \\
    &= 
    \left|\langle \nu_\beta|
    \exp\left(-i \frac{\Delta \tilde m^2_{j1}}{2E} L_2\right)
    \left(\sum_{j} |\tilde\nu_{j}\rangle\langle\tilde\nu_{j}|\right)
    \left(\sum_{\gamma} |\nu_{\gamma}\rangle\langle\nu_{\gamma}|\right)
    \right. \\  &\left.  \qquad\; \times
    \exp\left(-i \frac{\Delta m^2_{k1}}{2E} L_1\right)
    \left(\sum_{k} |\nu_{k}\rangle\langle\nu_{k}|\right)
    |\nu_\alpha\rangle\right|^2
    \\
    &= 
    \left|\sum_\gamma \sum_{j,k}\, 
    \tilde U_{\beta j}\tilde U_{\gamma j}^* U_{\gamma k} U_{\alpha k}^*\,
    \exp\left(-i \frac{\Delta m^2_{k1} L_1 + \Delta \tilde m^2_{j1} L_2}{2E} \right)
   \right|^2
    \,.
\end{split}
\label{eq:P2}
\end{equation}
It is clear that oscillation probabilities depend on the point of production of the (active) $\alpha$-flavor neutrino, located at a distance $L= L_1+L_2$ from the ND. As discussed in the introduction, our main goal is to quantitatively assess the impact of this dependence in the DUNE setup.

Event rates at the FD, on the other hand, are expected to be sensitive only to the average matter density~\cite{Kelly:2018kmb}, taken here to be $\rho\simeq 2.6\text{ g cm}^{-3}$. Furthermore, given its large distance to the FD, the source can be approximated as pointlike for the computation of FD events.

For both ND and FD event rates, a low-pass filter is applied at the probability level to appropriately average out unresolvable fast oscillations produced by large sterile mass-squared differences.
Averaging the oscillation probabilities over a Gaussian energy distribution with standard deviation $\sigma_E$ (see also section 7.6 of~\cite{Giunti:2007ry}), we find%
\footnote{This can be seen as a direct generalization of the implementation described in~\cite{Huber:2020}.
The averaging can be carried out by assuming that $1/E$ is normally distributed, which is a good approximation for sufficiently peaked distributions, see appendix C of~\cite{Coloma:2021uhq}.}
\begin{equation}
\begin{split}
    \langle P_{\alpha\beta}^\text{SBL}(L_i,E)\rangle &= 
    \sum_{j,k} \sum_{j',k'} \sum_{\gamma,\gamma'}
    \tilde U_{\beta j}   \tilde U^*_{\gamma j}\,\,
    \tilde U_{\gamma'j'} \tilde U^*_{\beta j'}\,\,
    U_{\gamma k}U^*_{\alpha k}\,\,
    U_{\alpha k'} U^*_{\gamma'k'}
    \\ &\,\times
    \exp\left(-i \frac{\Delta m^2_{kk'} L_1 + \Delta \tilde m^2_{jj'} L_2}{2E} \right)
    \exp\left[-\frac{\sigma_E^2}{2 E^2}
    \left(\frac{\Delta m^2_{kk'} L_1 + \Delta \tilde m^2_{jj'} L_2}{2E} \right)^2\right]
    \,,
\end{split}
\label{eq:PfilterND}
\end{equation}
where the rightmost exponential factor is responsible for phasing out the high frequencies.

Eq.~\eqref{eq:PfilterND} is appropriate for the simulation of ND events, while for the FD, one has
\begin{equation}
\begin{split}
    \langle P_{\alpha\beta}^\text{LBL}(L,E)\rangle &= 
    \sum_{j,j'} \tilde U_{\beta j} \tilde U^*_{\alpha j} \, \tilde U_{\alpha j'}  \tilde U^*_{\beta j'}\, \exp\left(-i\frac{\Delta \tilde m^2_{j j'} L}{2E}\right)\exp\left[-\frac{\sigma^2_E}{2E^2}\left(\frac{\Delta \tilde m^2_{j j'}L}{2E}\right)^2\right]
\end{split}
\label{eq:PfilterFD}
\end{equation}
in the limit of a single baseline $L$ in matter. 
Note that this filter is only considered due to the General Long Baseline Experiment Simulator (GLoBES) setup and sampling constraints. It should not overshadow the detector energy resolution (see~\cref{sec:chi2} for details on the chosen value). In a different setup, it could be exchanged by the increase in the precision of energy integrals --- although with no expected improvement in the accuracy of the results, given the limitation from the detector resolution.

We further employ analytical expressions for the matter-dependent quantities $\Delta \tilde m^2_{ij}$ and $\tilde U_{\alpha i}\tilde U_{\beta i}^*$, making use of the exact diagonalization of the matter Hamiltonian performed in~\cite{Zhang:2006yq,Li:2018ezt}.%
\footnote{Approximate results based on these works have been derived in~\cite{Yue:2019qat}.}
The relevant results are given in \cref{app:diag}, in
\cref{eq:mtNO,eq:mtIO,eq:UtUts,eq:FC}.
We have implemented the corresponding analytic probability engine within the GLoBES simulation software~\cite{Huber:2004ka,Huber:2007ji}. The relevant code is provided in the ancillary files \texttt{exact\_4nu.c} and \texttt{exact\_4nu.h}~\cite{Supplemental}.

Accounting for matter effects is crucial for an accurate estimation of FD event rates. For ND event rates, however, the level of detail contained in \cref{eq:PfilterND} turns out to be excessive in practice. As we numerically verify in what follows, it will be enough to consider the approximate result,
\begin{equation}
\begin{split}
    \langle P_{\alpha\beta}^\text{SBL}(L,E)\rangle &\simeq
    \delta_{\alpha\beta} - 2|U_{\alpha 4}|^2\left(\delta_{\alpha\beta} - |U_{\beta 4}|^2\right)\\
    &\times
    \Bigg\{1-\cos\left(\frac{\Delta m^2_{41} L}{2E} \right)
    \exp\left[-\frac{\sigma_E^2}{2 E^2}
    \left(\frac{\Delta m^2_{41} L }{2E} \right)^2\right]\Bigg\}
   \,,
\end{split}
\label{eq:SBL}
\end{equation}
obtained in the limit of vanishing matter density and vanishing standard-neutrino mass-squared differences (see \cref{app:approx} for a derivation), where $L=L_1+L_2$ is the event-dependent baseline for sterile oscillations. 

In the next section, we make use of \cref{eq:PfilterFD,eq:SBL} to assess the sterile exclusion reach of DUNE. We contrast the usual scenario where the event dependence of the baseline is neglected, i.e.,~where $P_{\alpha\beta}^\text{SBL}$ depends on a fixed baseline $L=574$ m, with the more accurate scenario where information on the neutrino production point is incorporated in the event rate estimation.

\section{A sterile neutrino in DUNE}
\label{sec:dune}

\subsection{Simulation details}
\label{sec:simulation}

The collisions of the primary proton beam on the graphite target and the resulting neutrino beam downstream from the hadron absorber have been simulated using version \texttt{v3r5p7} of the neutrino beamline simulator \texttt{g4lbne}~\cite{g4lbne}, built against version \texttt{4.10.3.p03} of Geant4~\cite{Agostinelli:2002hh,Allison:2006ve}. We have assumed a 1.2 MW, 120 GeV proton beam ($1.1\times 10^{21}$ POT/y) and a cylindrical target, 1.5 m long and 1.6 cm in diameter, followed by a system of three magnetic focusing horns, operated at a current of $\pm 300\text{ kA}$ for the FHC and RHC modes, respectively.

The \texttt{g4lbne} code allows one to compute unoscillated fluxes at the ND and the FD, given the positions of the detectors. The flux files are passed to the GLoBES software, which expects as input a single baseline $L$ for each detector (or ``experiment,'' in GLoBES terminology), as it assumes the source to be pointlike. Since the distance of the target to the ND is of the order of the decay volume length, the point-source approximation is not valid and source-volume geometry effects need to be taken into account --- as they indeed are, even in 3+0 analyses. A routinely-used weighting procedure is enough to circumvent the GLoBES point-source restriction and encode these effects into a single ND flux file.
This file, however, contains only the total unoscillated flux for each energy bin. Without further action, the information on the production point of each neutrino is lost to the subsequent stages of the simulation and an accurate simulation of the 3+1 scenario is impossible.

\begin{figure}[t]
\centering
\includegraphics[width=0.43\textwidth]{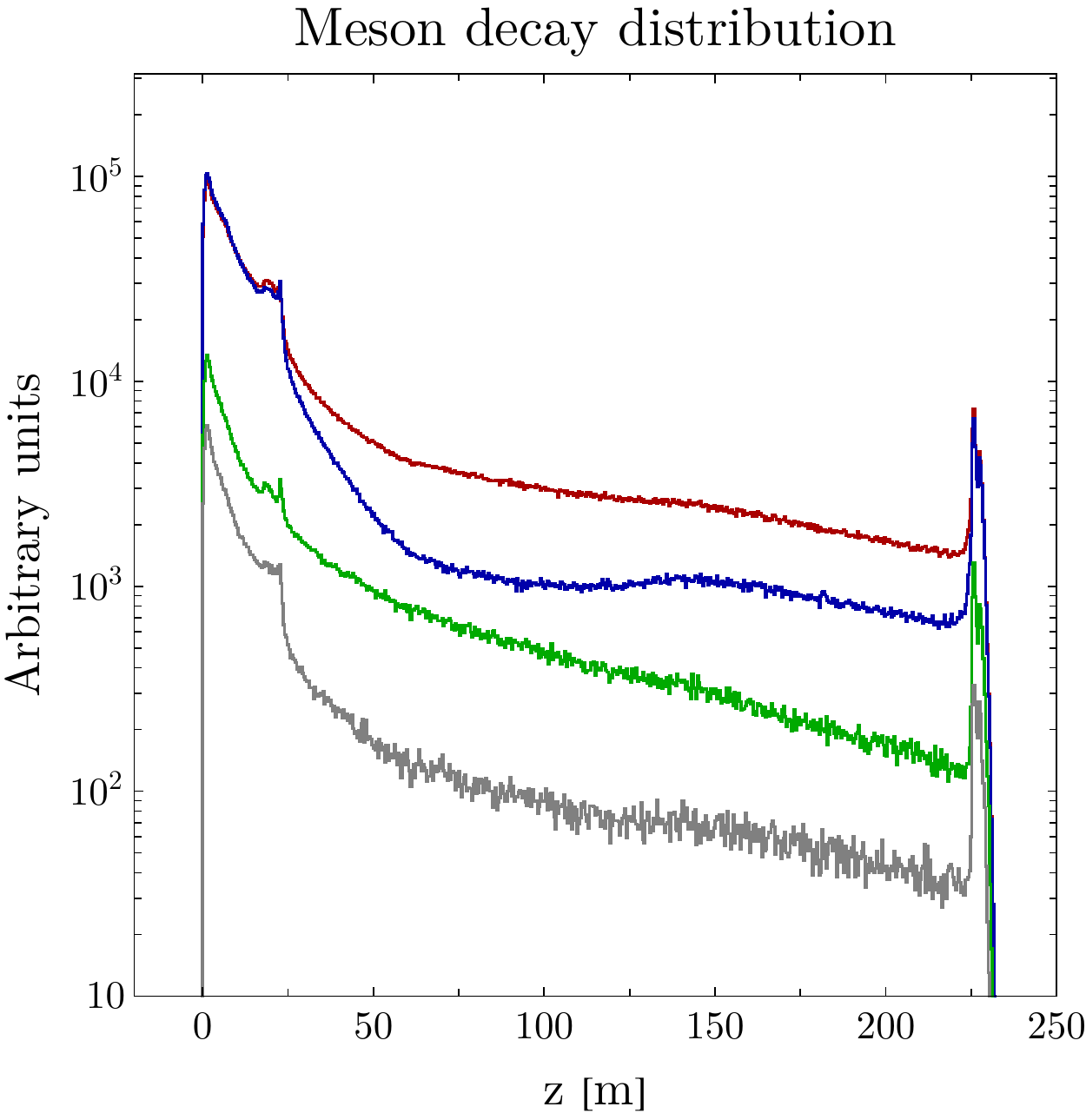}\quad
\includegraphics[width=0.51\textwidth]{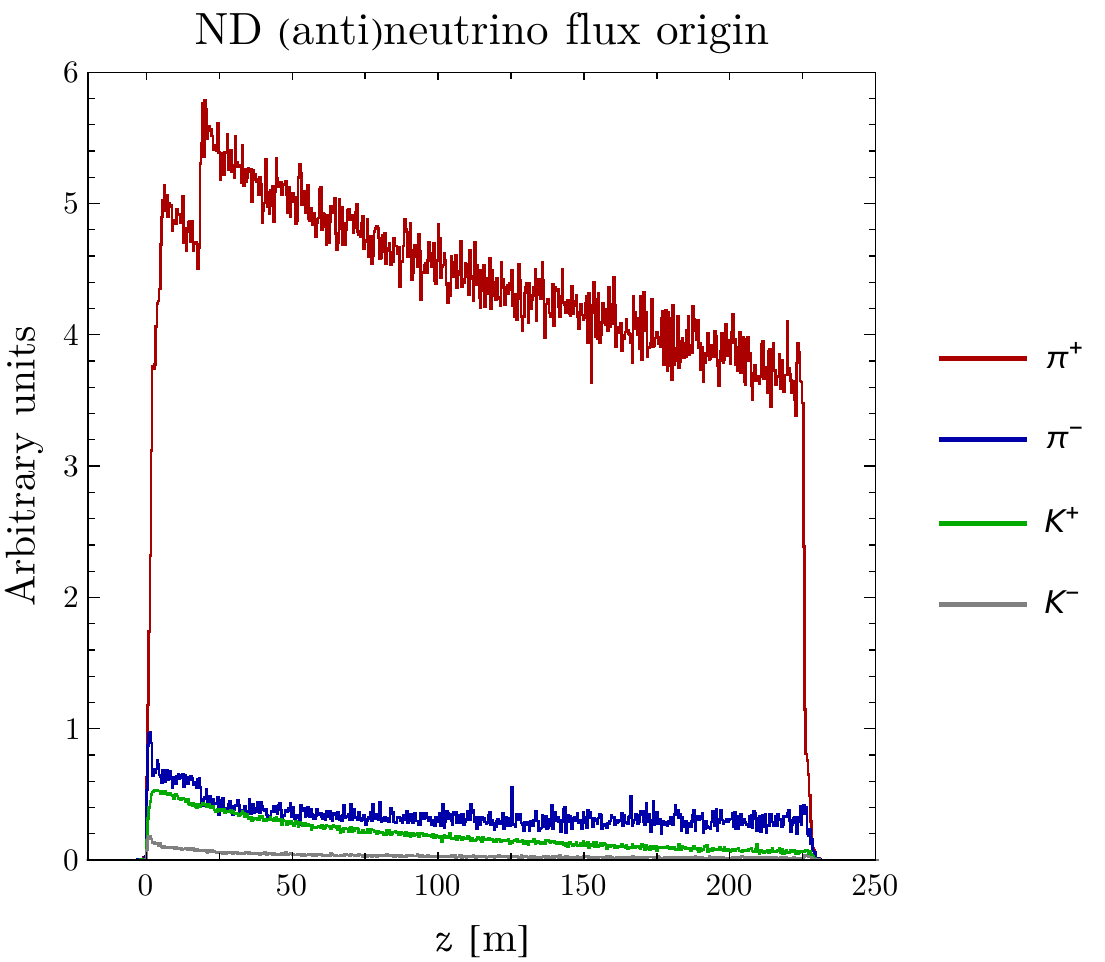}
\caption{Left panel: the meson decay distribution in FHC mode as a function of distance from the target (cf.~\cref{fig:setup}). Right panel: the origin of the corresponding neutrinos and antineutrinos reaching the ND.
}
\label{fig:dist}
\end{figure}

In \Cref{fig:dist} we show the distribution of meson decays along the decay volume in FHC mode (left panel), as well as the neutrino flux effectively reaching the ND as a function of the location of the parent meson decay, obtained following the weighting procedure (right panel).
This information on the distance traveled by the neutrinos may only enter GLoBES via i) the flux files produced by \texttt{g4lbne} or ii) the baseline $L$, which is used by the probability engine.
To incorporate into GLoBES the desired baseline dependence of oscillation probabilities for ND event rates in the 3+1 scenario, we conceptually divide the decay volume into sections and associate each to a different GLoBES point-source ``experiment'', with its own fixed (average) baseline $L$.
Thus, information on the distance traveled by the neutrinos enters the simulation at the level of the probability engine and via a flux normalization factor.%
\footnote{Within AEDL and supporting files, the variable \texttt{@norm} (see appendix C of~\cite{Huber:2020}) usually includes a factor of baseline length squared $L^2$, canceling a factor of $L^{-2}$ in the computation of event rates.}
The flux arriving at the ND from each section is then passed to GLoBES as the flux of an independent ``experiment''. Of course, these do not represent actual independent experiments or detectors, as there is no way of discriminating the neutrino section of origin at detection.
In this setup, one must therefore be careful to appropriately modify internal GLoBES functions, especially $\chi^2$ functions (see \cref{sec:chi2}), in order to sum over all the virtual sections when computing ND event rates.

In the GLoBES simulation, we have taken the ND and FD to correspond to fiducial liquid argon masses of 50 ton and 40 kt respectively, following~\cite{DUNE:2021tad}. We have assumed 3.5 years of operation in FHC mode, as well as 3.5 years for the RHC mode. The ND is taken to be a scaled-down version of the FD and GLoBES AEDL files have been adapted from those in ref.~\cite{DUNE:2021cuw}. We make use of the efficiencies, the smearing matrices, and the data on CC and NC neutrino interactions in argon from~\cite{DUNE:2021cuw}. The latter have been obtained using version \texttt{2.8.4} of the GENIE event generator~\cite{Andreopoulos:2009rq}.
Downstream from the target, 20 sections are considered, each 2.5 m deep, followed by 10 additional ones, each 18 m deep, for a total of 30 sections covering a distance of 230 m.
In order to collect enough statistics for each section, we have simulated $10^7$ proton events within \texttt{g4lbne} for each mode of operation, FHC and RHC.

\subsection{Chi-squared analysis}
\label{sec:chi2}

We are interested in producing exclusion plots in the $\left(\sin^2\theta_{14},\Delta m^2_{41}\right)$ and $\left(\sin^2\theta_{24},\Delta m^2_{41}\right)$ planes, comparing the single $L = 574$ m baseline computation with the more realistic one where the decay volume is sectioned, as described above.
Given DUNE's capabilities and following ref.~\cite{DUNE:2021cuw}, we consider four ``rules'' altogether, corresponding to CC events associated with
\begin{enumerate}
\item $\nu_e + \bar\nu_e$ appearance in FHC mode,
\item $\nu_e + \bar\nu_e$ appearance in RHC mode,
\item $\nu_\mu + \bar\nu_\mu$ disappearance in FHC mode, and
\item $\nu_\mu + \bar\nu_\mu$ disappearance in RHC mode.
\end{enumerate}
A ``rule,'' as defined in the GLoBES language (see~\cite{Huber:2020}), encompasses a number of signal and background channels and their associated systematic uncertainties. Each rule provides a separate contribution to the $\chi^2$. Taken together, the rules constitute the final link between the event computation and the statistical analysis.

Recall that in FHC and RHC modes, positive and negative meson decays along the decay pipe are responsible for the main component of the neutrino or antineutrino flux --- the $\nu_{\mu}$ or $\bar\nu_{\mu}$ component, respectively.
The signal channels for $\nu_e+\bar\nu_e$ appearance then correspond to $\barparene$ events arising from the oscillation of the $\barparenm$ component of the beam.
The relevant channels for $\nu_\mu+\bar\nu_\mu$ disappearance are likewise connected to the survival of the $\barparenm$ beam component, allowing for possible oscillations.
Background channels, on the other hand, originate from beam contaminations and misidentifications, comprising events that have similar final state properties with respect to the signal ones.
Contaminations in FHC mode are due to the $\bar\nu_{\mu}$, $\nu_e$, and $\bar\nu_e$ beam components, whereas in RHC they arise from the $\nu_{\mu}$, $\nu_e$, and $\bar\nu_e$ components.
Note that the $\barparenm$ contamination channels are included as signal in the disappearance analyses, which group neutrino and antineutrino events for either mode (in contrast with, e.g.,~\cite{Das:2014fja}).
Instead, the $\barparene$ contamination channels are one of the major backgrounds to $\nu_e+\bar\nu_e$ appearance. The remaining background channels correspond to misidentifications. These arise from $\barparenm$ mistakenly accepted as $\barparene$ and from NC events wrongly classified as CC events. The latter may arise from any component of the flux, and, in particular, from the main $\nu_\mu$ and $\bar \nu_\mu$ components.%
\footnote{ \label{footnote:NC}
The distinction between signal and background is blurred when dealing with NC misidentification, since NC events are reduced in the presence of a sterile. In this case, one should avoid the \texttt{NOSC\_} GLoBES prefix and instead detail the active flavor channels in the AEDL file.}
If sterile neutrinos exist, then $\nu_e$ and $\bar\nu_e$ can appear at the ND as signal due to sterile-induced oscillations. In the 3+0 case, $\nu_e$ and $\bar\nu_e$ events at the ND can only originate from contaminations or misidentifications.

\begin{figure}[t]
\centering
\includegraphics[width=0.48\textwidth]{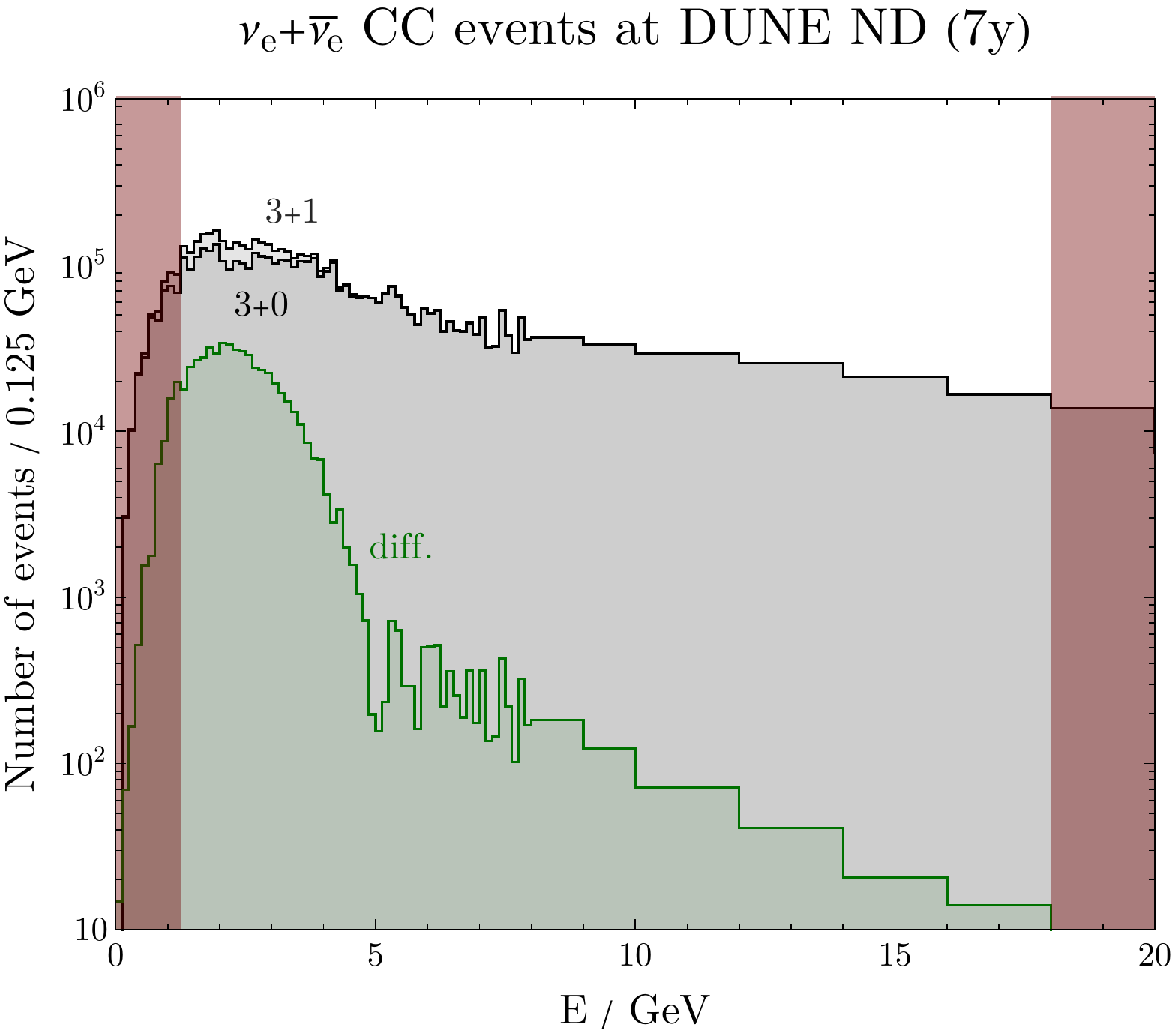}\quad
\includegraphics[width=0.48\textwidth]{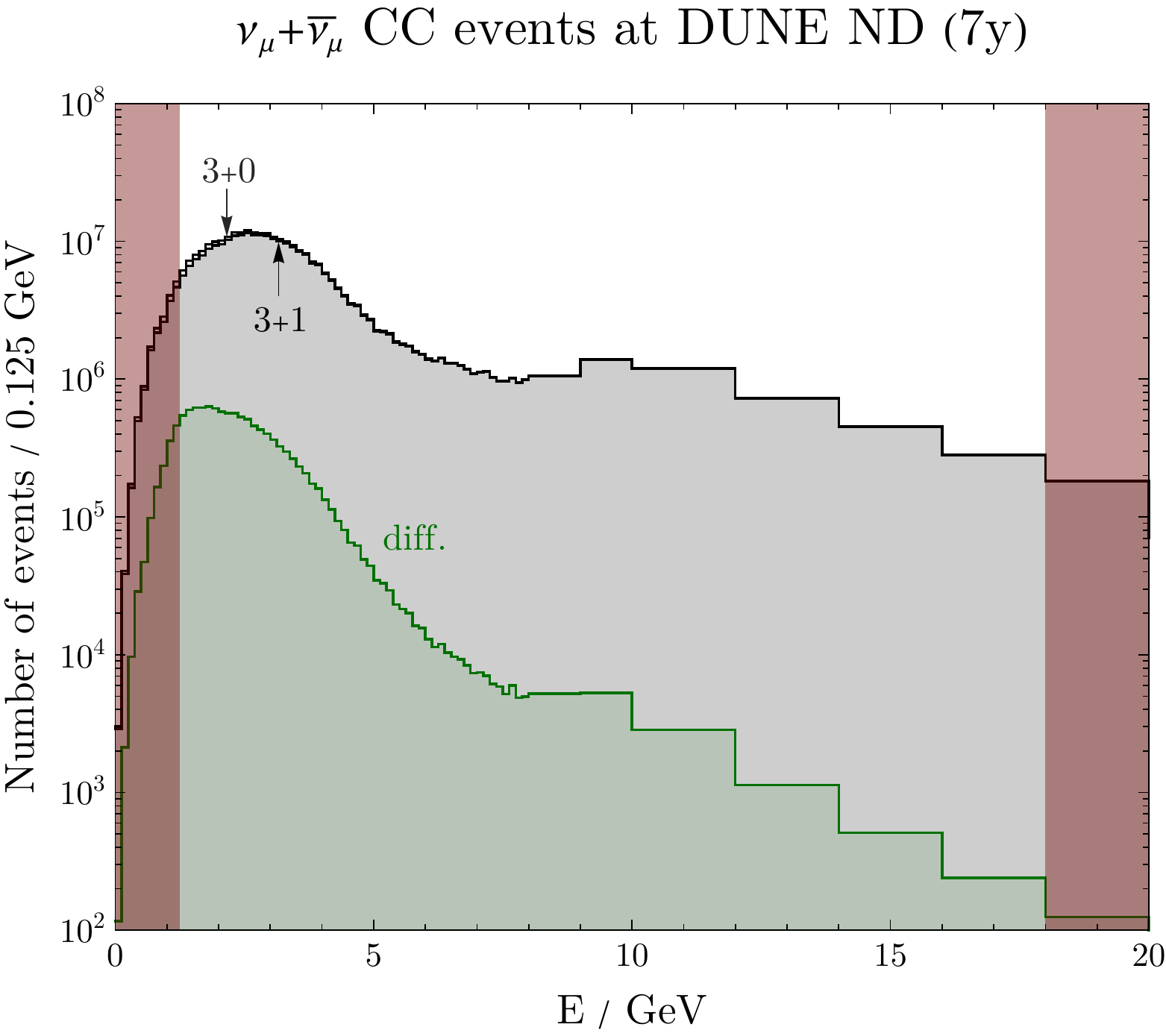}
\caption{CC event counts at the DUNE near detector after seven years of operation, as a function of the (anti)neutrino reconstructed energy, for $\Delta m^2_{41}=3\text{ eV}^2$, $\sin^2\theta_{14}=0.07$, $\sin^2\theta_{24}=0.03$, and $\sin^2\theta_{34}=\delta_{24}=\delta_{34}=0$.
Left panel: $\nu_e+\bar\nu_e$ appearance events in the 3+0 and 3+1 cases. The absolute values of the differences are also shown. Red vertical bands delimit the energy window considered in the analysis.
Right panel: the same for $\nu_{\mu}+\bar\nu_{\mu}$ disappearance events.}
\label{fig:events}
\end{figure}

For illustrative purposes, we show in \Cref{fig:events} the distribution of neutrino and antineutrino events at the ND, after the assumed seven years of operation. We compare the 3+0 case and the 3+1 case for a set of sterile parameters to which DUNE will be sensitive (cf.~results in the following section).
Differences between numbers of events are also displayed. The left- and right-hand panels correspond to $\nu_e + \bar\nu_e$ appearance and $\nu_\mu + \bar\nu_\mu$ disappearance, respectively.
Whereas $\barparene$ events in the 3+0 case come only from background channels through contaminations or misidentifications, in the 3+1 case there are extra such events originating from oscillations due to the sterile.%
\footnote{The NC misidentification background is also reduced (see \cref{footnote:NC}), and this reduction dominates over the extra signal events for $E \gtrsim 5$ GeV in the left panel of \cref{fig:events}.}
For $\barparenm$ (right panel) the number of events is reduced in the 3+1 case, as some $\barparenm$ will have oscillated to other flavors.
Here and in what follows, we take $\sigma_E = 0.125$ GeV for the low-pass filter and consider an energy window of $E\in [1.25,18.0]$ GeV containing 60 energy bins, as indicated in~\Cref{fig:events}. The lower limit on $E$ guarantees the validity of the Gaussian averaging leading to \cref{eq:PfilterND} (see also appendix C of~\cite{Coloma:2021uhq}). It can be checked that this value of $\sigma_E$ is below the energy resolution of the detectors, see e.g. Figure 1 of ref.~\cite{Chatterjee:2021wac}.

Once the binned event rates are obtained for each rule at both ND and FD, we proceed with the statistical analysis based on the minimization of the total $\chi^2$ function, defined as
\begin{equation}
\chi^2=\chi^2_\text{stat}(\omega,\omega_0,\zeta,\zeta')+\chi^2_\text{prior}(\omega,\omega_0)   +\sum_{k=1}^{26}\left(\frac{\zeta_k}{\sigma_k}\right)^2 
+\sum_{r=1}^{4}\sum_{i=1}^{60}\left(\frac{\zeta'_{r,i}}{\sigma'}\right)^2\,.
\label{eq:tot}
\end{equation}
The minimization is carried out over the parameter test values $\omega$ (see below) and over the normalization and shape nuisance parameters, $\zeta$ and $\zeta'$.
The index $k$ goes over the set of 26 independent normalization systematic errors. These are obtained from ref.~\cite{Abi:2020kei}, together with the associated prior uncertainties $\sigma_k$ and are collected in \Cref{tab:systematics}.
Energy bin-to-bin uncorrelated systematic errors (see also \Cref{tab:systematics}), which we consider at a later stage, are labeled by the indices $r$ and $i$. These indices refer to the four rules mentioned earlier and to the 60 energy bins, respectively. A common 5\% uncertainty ($\sigma' = 0.05$) is used for this set of 240 shape systematics.

\begin{table}[t]
\renewcommand{\arraystretch}{1.2}
\centering
\begin{tabular}{llcccc}
\toprule
Type & Description & $\sigma$ & Det. & Rules & Channels
\\
\midrule
\multirow{14}{*}{\shortstack[l]{Normalization\\[2mm] ($\zeta_k$)}}
& ND fiducial volume & $1\%$ & ND & all & all\\
& FD fiducial volume & $1\%$ & FD & all & all\\
& flux for FHC signal channels & 8\% & both & 1,\,3 & sig.\\
& flux for RHC signal channels & 8\% & both & 2,\,4 & sig.\\
& flux for FHC background channels & 15\% & both & 1,\,3 & bkg.\\
& flux for RHC background channels & 15\% & both & 2,\,4 & bkg.\\
& flux for FHC sig.~(ND/FD diff.) & 0.4\% & ND  & 1,\,3 & sig.\\
& flux for RHC sig.~(ND/FD diff.) & 0.4\% & ND  & 2,\,4 & sig.\\
& flux for FHC bkg.~(ND/FD diff.) & 2\% & ND  & 1,\,3 & bkg.\\
& flux for RHC bkg.~(ND/FD diff.) & 2\% & ND  & 2,\,4 & bkg.\\
& CC cross sections for $\barparena$ (6 syst.) & 15\% & both  & all & all $\barparena$\\
& CC xsec.~(ND/FD diff., 6 syst.) & 2\% & ND  & all & all $\barparena$ \\
& NC cross sections for $\barparenn$ (2 syst.) & 25\% & both  & all & NC bkg.\\
& NC xsec.~(ND/FD diff, 2 syst.) & 2\% & ND  & all & NC bkg.\\
\midrule
\multirow{4}{*}{\shortstack[l]{Shape ($\zeta'_{r,i}$)\\ (bin-to-bin\\ uncorrelated)}}
& FHC $\barparene$ app.~dominant (60 syst.) & \multirow{4}{*}{5\%} & \multirow{4}{*}{both} & 1 & bkg. \\
& RHC $\barparene$ app.~dominant (60 syst.)  & & & 2 & bkg. \\
& FHC $\barparenm$ dis.~dominant (60 syst.) & & & 3 & sig. \\
& RHC $\barparenm$ dis.~dominant (60 syst.) & & & 4 & sig. \\
\bottomrule
\end{tabular}
  \caption{Summary of the systematic uncertainties considered in the analysis. Normalization systematics are obtained from ref.~\cite{Abi:2020kei}, ($\alpha = e,\mu,\tau$). Systematics are correlated across the indicated detectors, rules, and channels.}
  \label{tab:systematics}
\end{table}

The first term in \mbox{\cref{eq:tot}} compares the event rates $T^d_{r,i}(\omega,\zeta,\zeta')$ obtained in the ``test'' 3+1 scenario with the event rates $O^d_{r,i}(\omega_0)$ obtained in the ``true'' or ``observed'' 3+0 scenario. It reads
\begin{equation}
\chi^2_\text{stat}=2\sum_{d=1}^{2}\sum_{r=1}^{4}\sum_{i=1}^{60}\left[T^d_{r,i}-O^d_{r,i}\left(1-\ln\frac{O^d_{r,i}}{T^d_{r,i}}\right)\right]\,.
\label{eq:stat}
\end{equation}
Here, the index $d$ refers to the detector (ND or FD). It is essential to sum the contributions of all the different decay volume sections (interpreted as GLoBES ``experiments'') in computing $T^\text{ND}_{r,i}$ and $O^\text{ND}_{r,i}$, before these quantities are introduced into $\chi^2_\text{stat}$. This is accomplished by a nontrivial modification of GLoBES internal functions, mainly at the level of the source files \texttt{glb\_minimize.c} and \texttt{glb\_sys.c}, so that one effectively works with two experiments/detectors and the two corresponding nonzero matter densities.

The 3+1 event rates $T^d_{r,i}$ depend on the test values $\omega$ and additionally on the relevant nuisance parameters $\zeta$ and $\zeta'$.
The vector $\omega$ contains the three mass-squared differences, the six mixing angles, the three CPV phases, and the nonzero matter densities, $\rho_\text{ND}$ and $\rho_\text{FD}$.
The value of the nonstandard mass-squared difference $\Delta m^2_{41}$ is fixed during the $\chi^2$ minimization and, depending on the analysis, we also fix either $\theta_{14}$ or $\theta_{24}$ in order to obtain exclusion plots in the corresponding planes.
For simplicity, we focus on a neutrino spectrum with normal ordering and also fix $\delta_{13}$ to a best-fit value $\delta_{13}^\text{b.f.} = 1.28\,\pi$~\cite{Capozzi:2017ipn}. We marginalize over the remaining 11 parameters contained in $\omega$, i.e.~they are free to vary in the minimization.
The 3+0 event rates $O^d_{r,i}$ are evaluated using the parameter vector $\omega_0$, containing the central values of the standard oscillation parameters and densities, with sterile mixing angles being set to zero.
We have
\begin{equation}
T^d_{r,i} = \sum_{\text{channel } c} N^d_{r,c,i} (\omega)
\left(1+\sum_l\zeta^d_{r,c,l}+ \zeta'_{r,c,i}\right)\,, \quad\,\,\,
O^d_{r,i} = \sum_{\text{channel } c} N^d_{r,c,i} (\omega_0)
\,.
\end{equation}
Here, the $N^d_{r,c,i}$ are the event rates for a given detector $d$, rule $r$, channel $c$, and energy bin $i$. The sum in $c$ sequentially goes over the relevant signal and background channels for each rule, while the index $l$ goes over all normalization systematics $\zeta$ contributing to a given channel.
Nuisance parameters are correlated across rules, channels, and detectors, as indicated in \Cref{tab:systematics}.%
\footnote{Note that to obtain properly correlated systematics one may need to resort to explicit calls of the as yet undocumented function \texttt{glbCorrelateSys}, which can be found in the GLoBES source file \texttt{glb\_multiex.c}.}
To avoid a proliferation of shape systematics, these are introduced only for the expected dominant component of each rule, namely for the background events in $\barparene$ appearance rules and for signal events in the $\barparenm$ disappearance rules. Thus, nonzero shape systematics $\zeta'$ are effectively labeled only by the indices $r$ and $i$, as previously indicated.

The second term in \cref{eq:tot} is the so-called prior term and includes information on measured neutrino parameters and their uncertainties, obtained from ref.~\cite{Capozzi:2017ipn}, as well as a possible 5\% uncertainty on the Earth density parameters. It reads
\begin{equation}
\chi^2_\text{prior}=\sum_{j=1}^{7}\left(\frac{\omega_j-(\omega_0)_j}{\sigma(\omega_j)}\right)^2\,,
\label{eq:prior}
\end{equation}
where the sum extends only over the seven measured parameters $\omega_j$ with central values $(\omega_0)_j$ and standard deviations $\sigma(\omega_j)$, i.e.,~over two mass-squared differences, three mixing angles and two average densities.
Finally, the last terms in \cref{eq:tot} correspond to penalty terms, appropriately controlling the magnitude of each nuisance parameter.

\subsection{Results}
\label{sec:results}

\begin{figure}[t]
\centering
\includegraphics[width=0.48\textwidth]{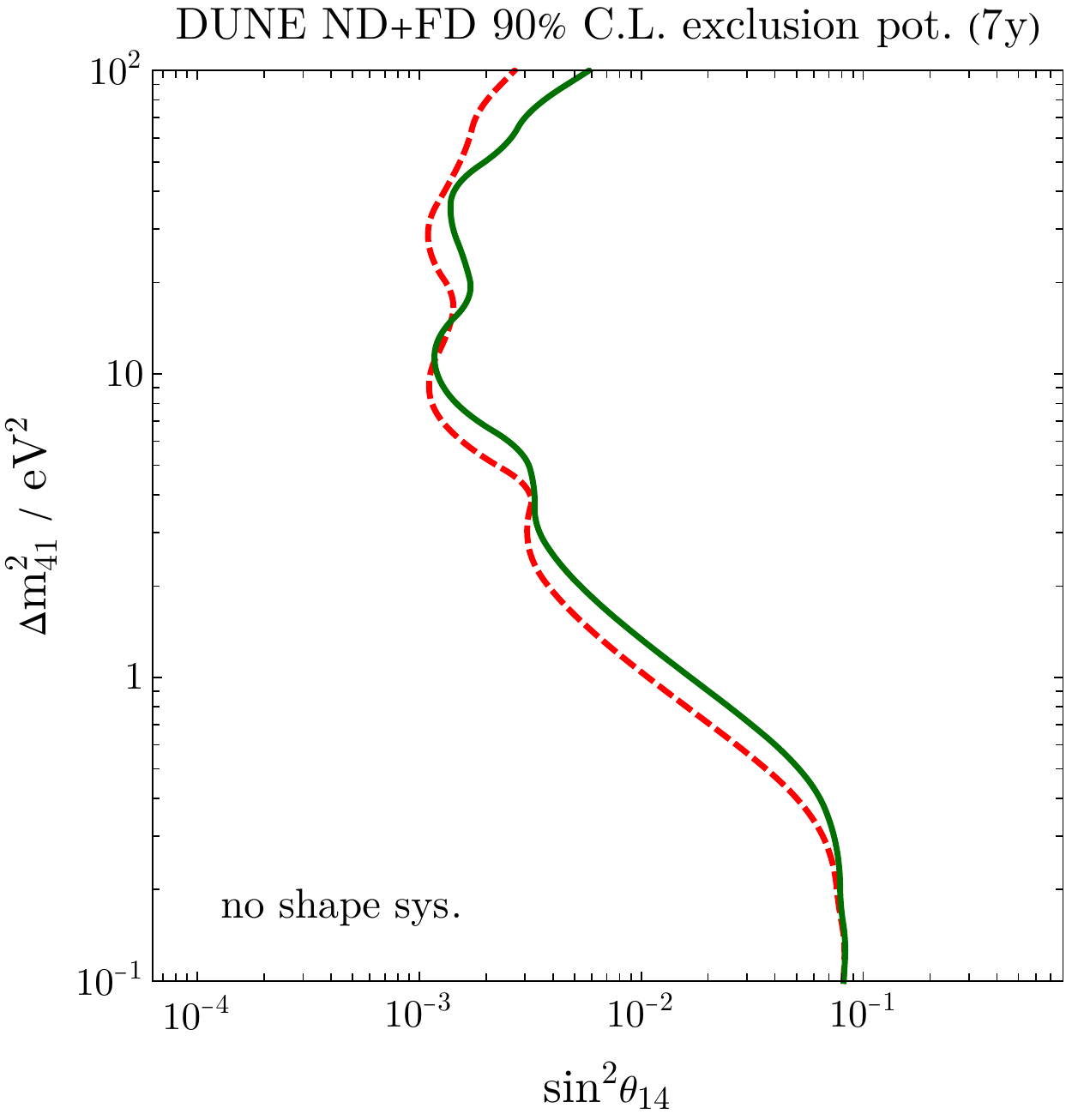}\quad
\includegraphics[width=0.48\textwidth]{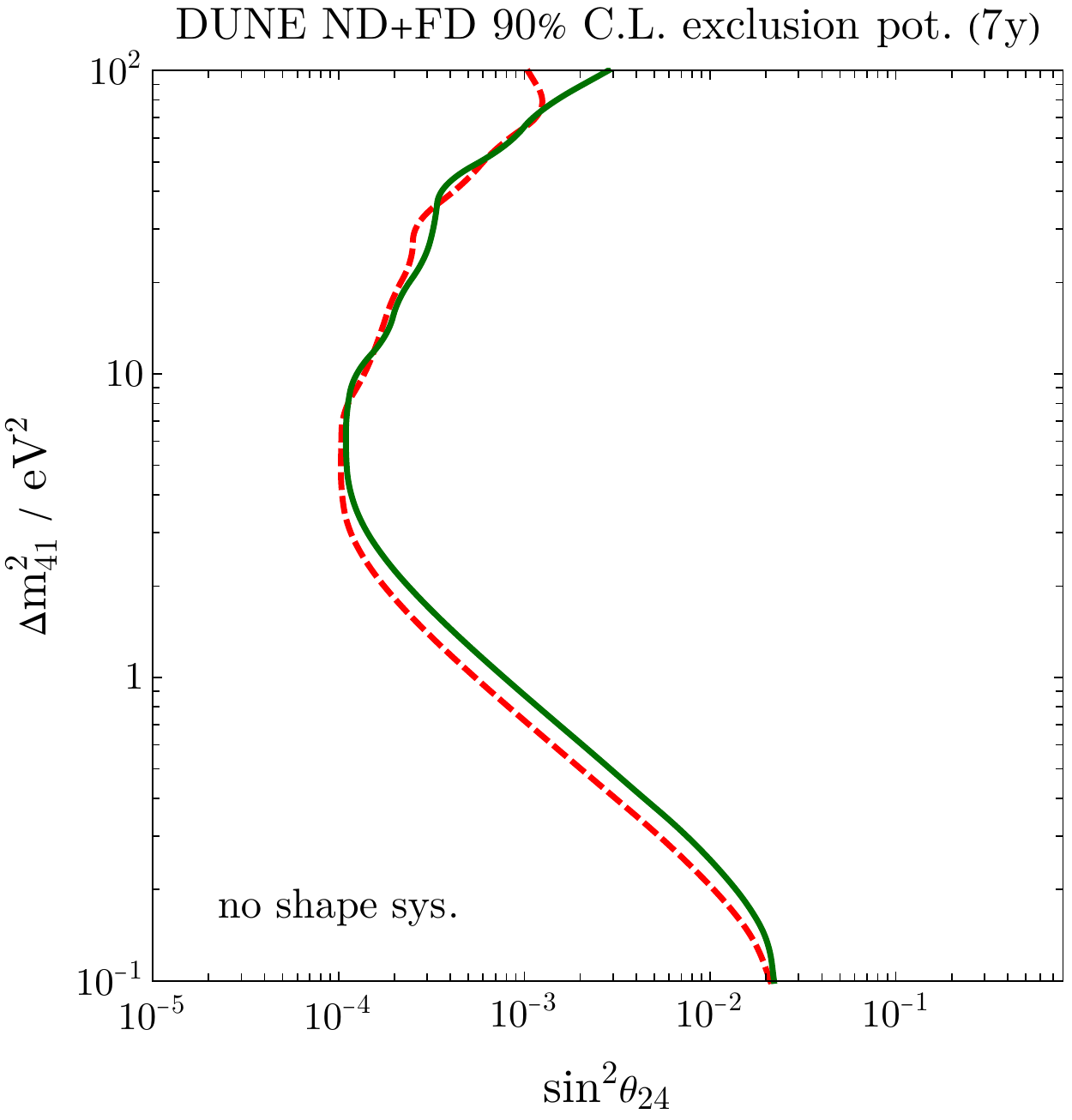}
\caption{
The sterile exclusion potential of the combined DUNE near and far detector CC analyses at 90\% C.L.~after seven years of operation. The dashed red curves are obtained assuming a common baseline of $L = 574$ m for near detector oscillations, while for the solid green curves the described smeared-source effect is taken into account.}
\label{fig:nosys}
\end{figure}

In this section we present the results of minimizing the $\chi^2$ function of \mbox{\cref{eq:tot}} independently in the $\left(\sin^2\theta_{14},\Delta m^2_{41}\right)$ and $\left(\sin^2\theta_{24},\Delta m^2_{41}\right)$ planes. At a first stage, we switch off shape systematics ($\zeta' = 0$) and vary the $11+26 = 37$ free parameters contained in $\omega$ and $\zeta$.
Our results are shown in \Cref{fig:nosys}, where the usual computation of test events using a single baseline $L= 574$ m at the oscillation probability level (dashed red lines) is compared to the more realistic one, with a different $L$ for each section of the decay volume (solid green lines). For each plane, the exclusion curves correspond to the locus of all points for which $\chi^2_\text{min}=4.61$ (90\% C.L.~for 2 d.o.f.).
It is seen that for most of the values of the sterile mass-squared difference $\Delta m^2_{41}$, the inclusion of the information on the neutrino production zone decreases the sensitivity of DUNE to the sterile mixing angles $\theta_{14}$ and $\theta_{24}$.
This is manifested in the fact that the green curve in both graphs is distorted and displaced to the right relative to the red one. This effect is present in the case of $\theta_{14}$ for $\Delta m^2_{41}\gtrsim 0.3 \text{ eV}^2$, while it is more pronounced for $0.1 \lesssim \Delta m^2_{41}\lesssim 4 \text{ eV}^2$ in the case of $\theta_{24}$.

\begin{figure}[t]
\centering
\includegraphics[width=0.48\textwidth]{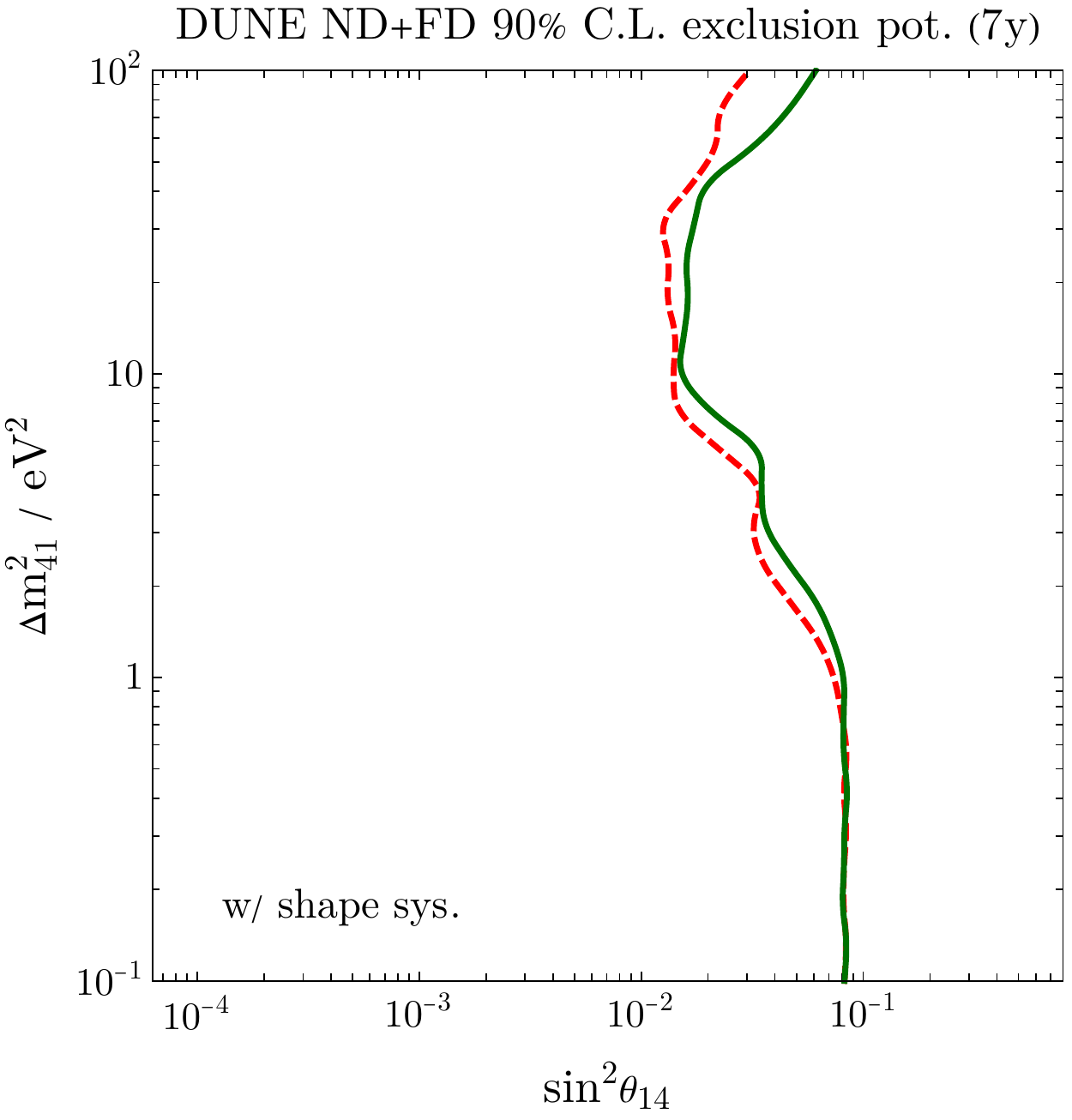}\quad
\includegraphics[width=0.48\textwidth]{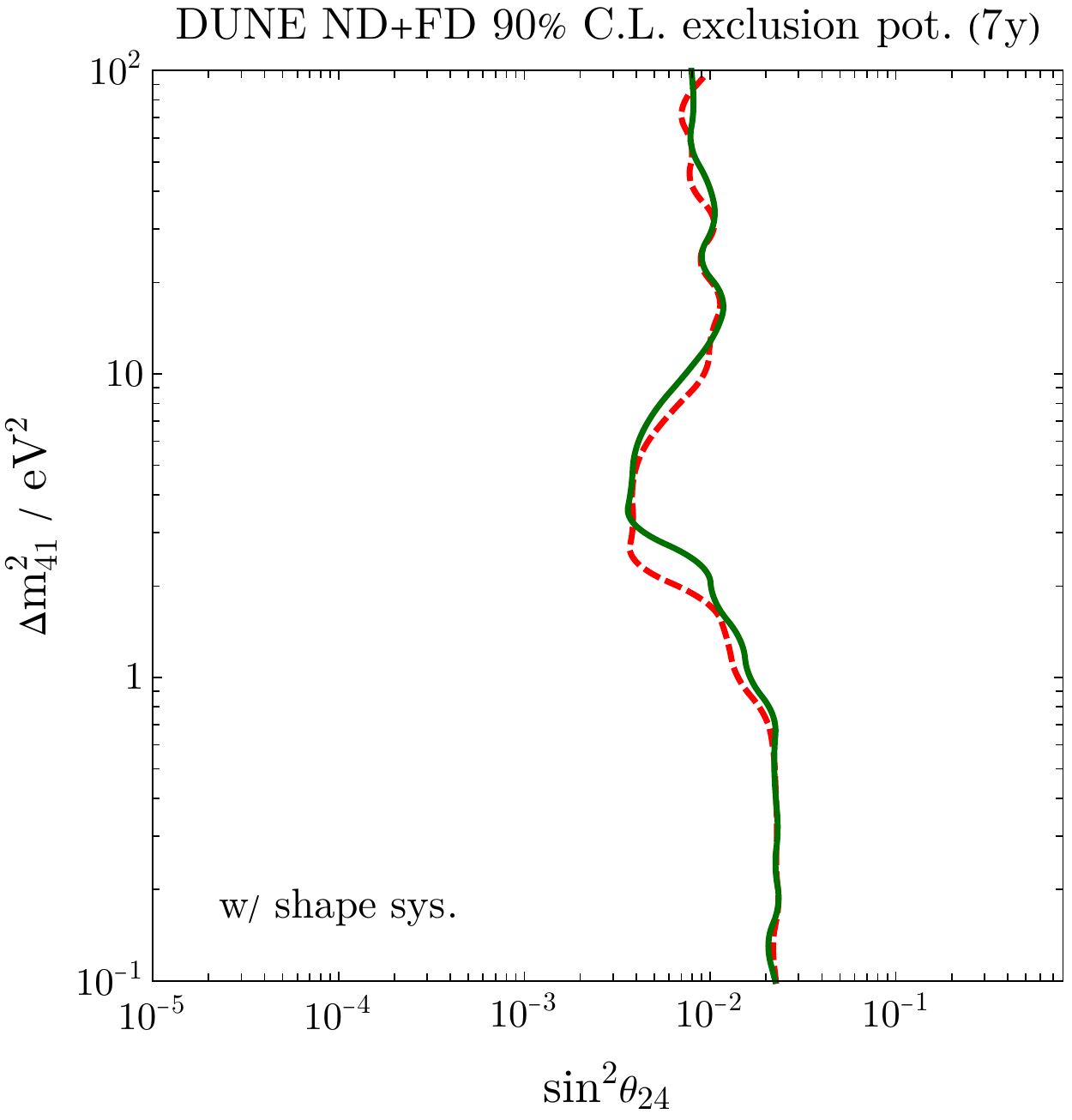}
\caption{The same as in \cref{fig:nosys}, now including additional $5\%$ energy bin-to-bin uncorrelated systematic errors.}
\label{fig:shape}
\end{figure}

We have repeated the minimization procedure including the bin-to-bin uncorrelated systematic errors $\zeta'$, which implies marginalizing over 240 additional parameters (for a total of 277). The result is shown in \Cref{fig:shape}.
There is a dramatic loss of sensitivity to sterile neutrino mixing parameters for both curves when the $\zeta'$ are switched on, as one may expect (cf.~ND-only analyses~\cite{DUNE:2021tad,Coloma:2021uhq} including shape systematics).
However, visible differences persist between the point-source (red) and smeared-source (green) exclusion curves even in this case. They indicate a decreased sensitivity that arises for $\Delta m^2_{41}\gtrsim 1 \text{ eV}^2$ in the $\theta_{14}$ plane and mainly for $\Delta m^2_{41} \sim \text{ few eV}^2$ in the $\theta_{24}$ plane.

\begin{figure}[t]
\centering
\includegraphics[width=0.48\textwidth]{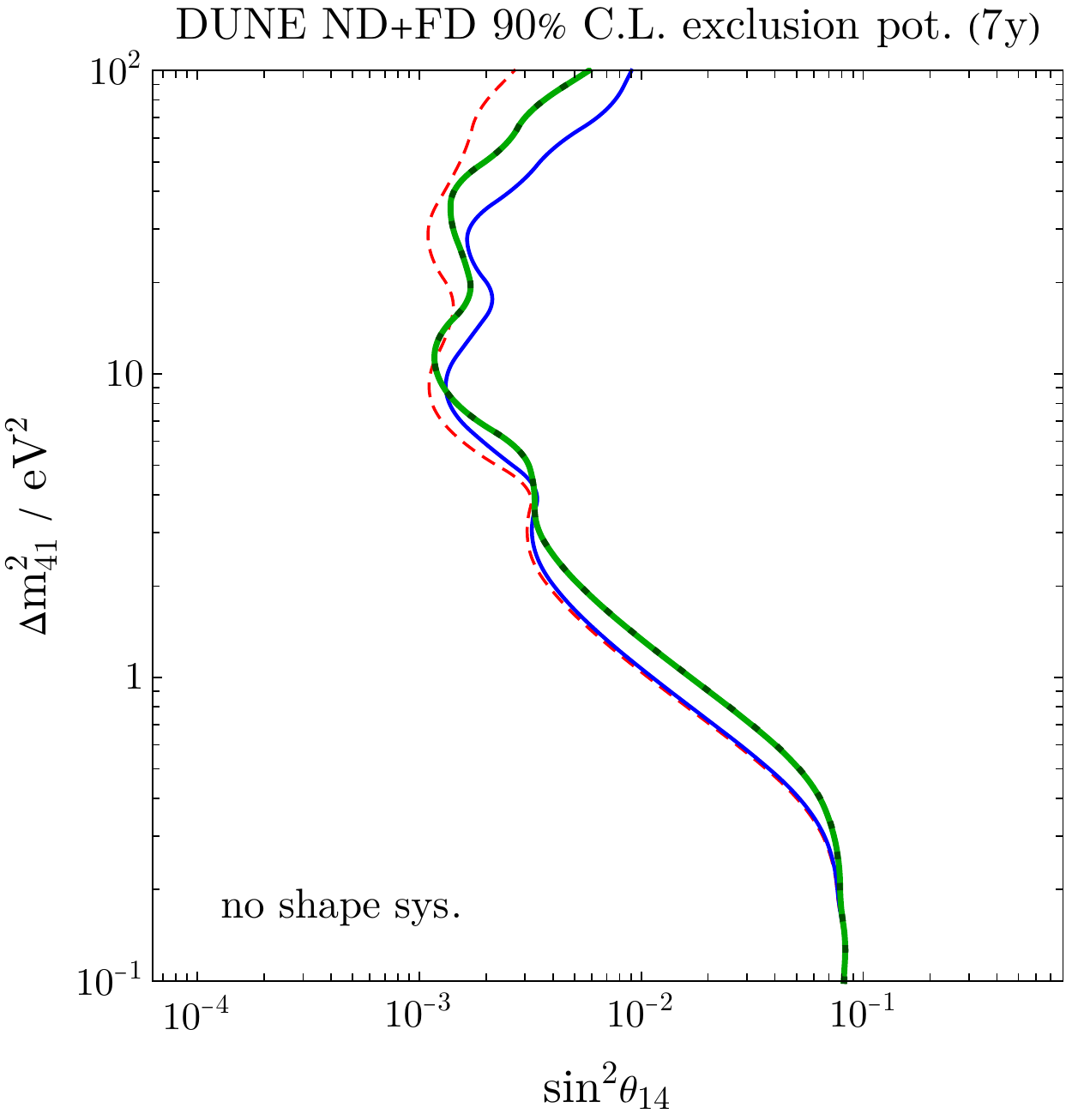}\quad
\includegraphics[width=0.48\textwidth]{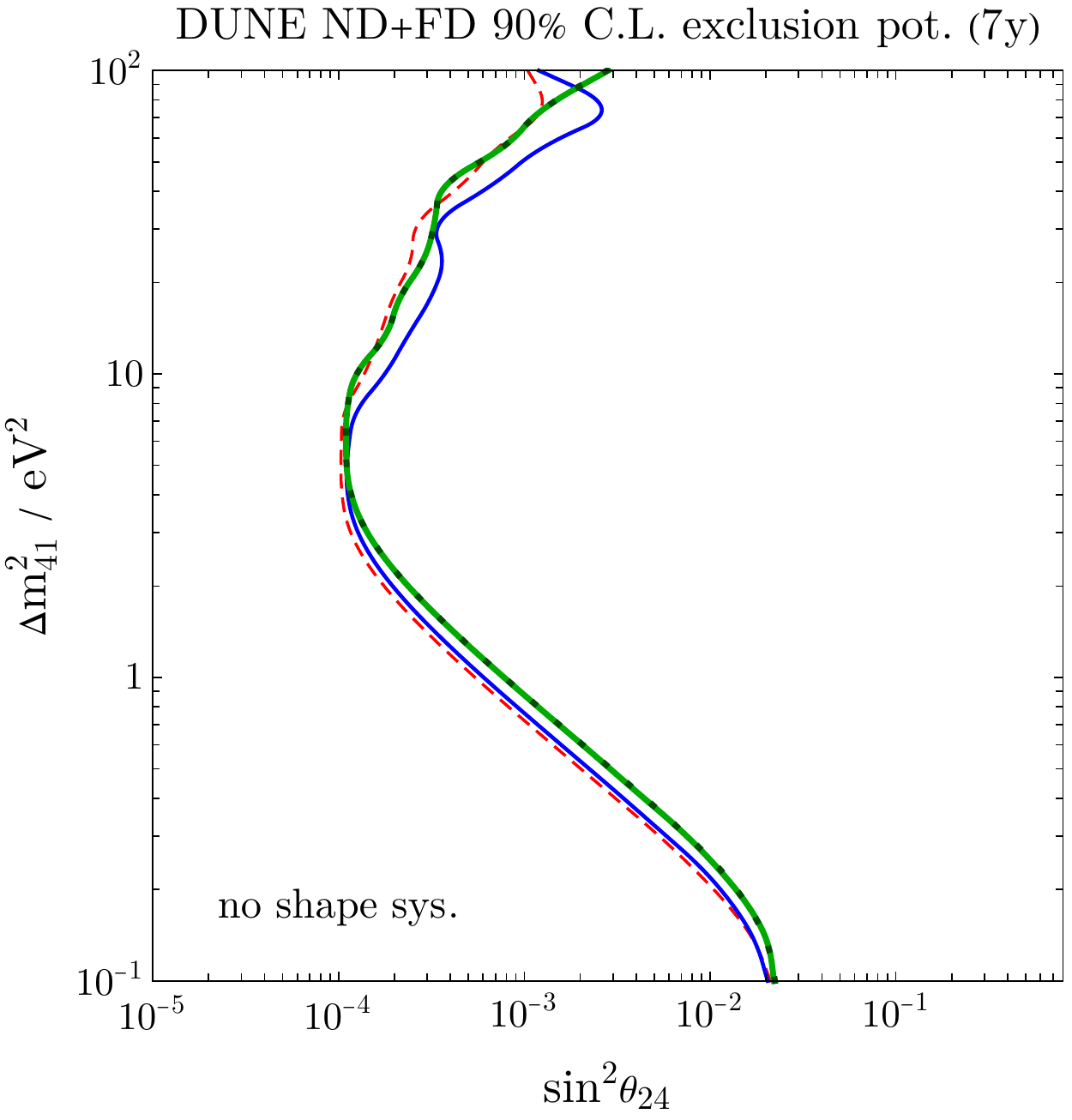}
\caption{
The same as in \cref{fig:nosys} (no shape systematics considered), with additional exclusion curves superimposed. The solid blue curves are obtained following the heuristic procedure described in ref.~\cite{Abi:2020kei}. Dotted dark green curves are obtained using \cref{eq:PfilterND}, which takes into account matter effects at the ND, and overlap with those obtained using the approximate SBL formula of \mbox{\cref{eq:SBL}} (solid green).
}
\label{fig:smearing}
\end{figure}

The possible relevance of taking into account the neutrino production point in the estimation of 3+1 ND events has previously been recognized in ref.~\cite{Abi:2020kei}. 
Given that oscillation probabilities depend on the baseline via the ratio $L/E$, a heuristic additional energy smearing has been used in~\cite{Abi:2020kei} to mimic the smeared-source effect (while keeping a single baseline), in place of a more precise computation like the one considered here. In particular, the GLoBES migration matrices for the DUNE ND have been multiplied by an extra smearing matrix, obtained from the integration of a Gaussian with a 20\% standard deviation in energy.
Reproducing this procedure, we compute and present the corresponding 90\% C.L.~exclusion curves for the case $\zeta'= 0$ in \Cref{fig:smearing} (blue lines), comparing them with the ones obtained previously.
From this figure, one can see that the heuristic procedure does not, in general, provide a valid approximation to the more realistic case. Overall, for both sterile mixing angles, the correction seems to be underestimated for $\Delta m^2_{41} \lesssim 3\text{ eV}^2$ and overestimated for $\Delta m^2_{41} \gtrsim 10\text{ eV}^2$.

As mentioned in \cref{sec:framework}, one may use the approximate result of \cref{eq:SBL} to assess the sterile exclusion potential of DUNE. We have numerically checked that this formula is sufficient for our purposes, i.e.~that taking into account matter effects at the ND using exact expressions~\cite{Li:2018ezt} and a two slab matter profile (with a nonzero density $\rho_\text{ND}$) via \cref{eq:PfilterND} does not produce noticeable differences with respect to the considered limit of vanishing matter density and vanishing standard-neutrino mass-squared differences. This is evidenced by the overlap of the solid and dotted green lines in \Cref{fig:smearing}.

To conclude this section, we comment on the behavior of the exclusion curves in the limits of small and large $\Delta m^2_{41}$. In both these limits, the probabilities $\langle P_{\alpha\beta}^\text{SBL}\rangle$ become independent of $L$, the effects of a smeared source disappear at the probability level, and the curves in \cref{fig:nosys,fig:shape} are expected to approach each other.
In fact, for sufficiently large $\Delta m^2_{41}$, oscillations should rapidly average out due to the presence of the low-pass filter exponential factor (see \mbox{\cref{eq:SBL}}).
On the other hand, in the limit of small $\Delta m^2_{41}$, the sterile oscillation length becomes much larger than the distance traveled by the neutrinos from their production point to the ND. 
Therefore, while sterile effects may be detected at the FD, no effect will be present at the ND, and event rates become independent of the point of neutrino production within the decay volume.
Whereas the curves do approach each other for small sterile mass-squared differences, it is not so in the large $\Delta m^2_{41}$ limit since a large energy $E$ in the low-pass filter exponential can counteract the largeness of $\Delta m^2_{41}$ for the considered energy window and range of sterile masses.
Note finally that the analyses presented here involve both detectors (ND and FD) and are thus able to provide bounds on sterile mixing for small values of $\Delta m^2_{41}$ (see also~\cite{Abi:2020kei}), unlike ND-only analyses~\cite{DUNE:2021tad,Coloma:2021uhq}.
Indeed, we find that the obtained exclusion curves for light sterile masses ($\Delta m^2_{41} \lesssim 0.8$ eV$^2$) are driven by FD event rates, which dominate the $\chi^2$ in this regime.
As the sterile oscillation lengths approach from above the scale of ND baselines,
the sensitivity reach becomes fully attributable to ND event rates (observed for $\Delta m^2_{41} \gtrsim 0.8$ eV$^2$).
Importantly, the differences observed between exclusion curves with the pointlike source approximation and without it (smeared source) remain even in the limit of no FD contribution to the $\chi^2$.

\section{Summary and Conclusions}
\label{sec:summary}

The possible existence of light sterile neutrinos has been hinted at by several experiments. The upcoming Deep Underground Neutrino Experiment is expected to play a clarifying role. The DUNE near detector, whose main purpose is to measure the unoscillated neutrino flux in a 3+0 scenario, can play a crucial role in constraining the 3+1 parameter space since the presence of a sterile neutrino may induce short-baseline oscillations of the active neutrinos. These oscillations depend on the baseline effectively traveled by the neutrino, which can vary considerably between events, given the length of the decay volume when compared to the distance from the target to the near detector.

In this work, we have analyzed the sterile exclusion potential of DUNE, taking into account both near and far detector event rates and shape uncertainties. Unlike previous studies, which consider a single sterile oscillation baseline when computing oscillation probabilities, we have sought to quantify the effect of the event-dependent baseline. 
We find that taking into account such information on the neutrino production point, in contrast to assuming a pointlike source, affects DUNE's sterile exclusion reach. In most of the parameter space, the inclusion of this smeared-source effect leads to a decrease in sensitivity (see \cref{fig:nosys}).
Furthermore, this effect is seen not to be appropriately modeled by an additional 20\% Gaussian energy smearing (see \cref{fig:smearing}). 
Finally, we have verified that differences between the pointlike source and smeared-source computations persist even if one takes into account energy bin-to-bin uncorrelated systematic errors (see \cref{fig:shape}), for which we have assumed a common 5\% standard deviation. 

Throughout our study, we have used the exact formulae of ref.~\cite{Li:2018ezt} (see~\cref{app:diag}) for the evaluation of the matter oscillation probabilities affecting far detector event rates. As for the near detector, we have verified that the approximate short-baseline formula of \cref{eq:SBL} and the exact one of \cref{eq:PfilterND}, which includes a two slab density profile, in practice, lead to the same results (see \cref{fig:smearing}).

\section*{Acknowledgements} 

We are indebted to Sampsa Vihonen for extensive support and insightful discussions. 
J.T.P further thanks M.~Orcinha for assistance with ROOT.
The work of J.T.P.~was supported by Fundação para a Ciência e a Tecnologia (FCT, Portugal) through the Projects No.~PTDC/FIS-PAR/29436/2017, 
  No.~CERN/FIS-PAR/0004/2019, No.~CERN/FIS-PAR/0008/2019,
  and CFTP-FCT Unit No.~UIDB/00777/2020 and No.~UIDP/00777/2020,
which are partially funded through Programa Operacional Ciência Tecnologia Inovação (POCTI) (Fundo Europeu de Desenvolvimento Regional (FEDER)), Programa Operacional Competitividade e Internacionalização (COMPETE), Quadro de Referência Estratégica Nacional (QREN) and European Union (EU). CFTP computing facilities were extensively used throughout the project.

\appendix
\section{Exact diagonalisation of the 3+1 matter Hamiltonian}
\label{app:diag}

\newcommand{\mt}[1]{{\Delta \tilde m^2_{#1}}}
\newcommand{\mh}[1]{{\Delta \hat m^2_{#1}}}
\newcommand{\m}[1]{{\Delta m^2_{#1}}}
\newcommand{\Ut}{{\tilde U}}
\newcommand{\A}[1]{\mathcal{A}_{#1}}
\newcommand{\Ah}[1]{{\hat{\mathcal{A}}_{#1}}}
\renewcommand{\aa}{\A{\alpha\alpha}}
\newcommand{\bb}{\A{\beta\beta}}
\newcommand{\haa}{\Ah{\alpha\alpha}}
\newcommand{\hbb}{\Ah{\beta\beta}}

\subsection{Mass-squared differences in matter}

Analytical expressions for the eigenvalues $\hat \Delta m^2_{i1}$ ($i=1,...,4$) of the matrix $2E H^\text{mat}$ have been obtained in ref.~\cite{Li:2018ezt} by explicitly solving the corresponding quartic characteristic equation. These eigenvalues read, in ascending order,
\begin{equation}
\begin{split}
&
- \frac{b}{4} - S - \frac{1}{2} \sqrt{-4S^2-2p+\frac{q}{S}}\,,\quad
- \frac{b}{4} - S + \frac{1}{2} \sqrt{-4S^2-2p+\frac{q}{S}}\, ,\\[2mm]
& - \frac{b}{4} + S - \frac{1}{2} \sqrt{-4S^2-2p-\frac{q}{S}}\,,\quad
- \frac{b}{4} + S + \frac{1}{2} \sqrt{-4S^2-2p-\frac{q}{S}} \,.
\end{split}
\end{equation}
The quantities $b$, $S$, $p$ and $q$ depend on $A_\text{CC}$, $A_\text{NC}$ and vacuum parameters. They read
\begin{equation}
  S = \frac{1}{2}\sqrt{- \frac{2}{3} p + \frac{2}{3} \sqrt{K_0} \cos \frac{\phi}{3}} \,, \qquad
  p = c - \frac{3}{8} b^2 \,, \qquad
  q = d - \frac{1}{2} b c + \frac{1}{8} b^3\,,
\end{equation}
where $K_0 =  c^2 - 3bd + 12 e $ and $\phi = \arccos \big({K_1}/{2 K_0^{3/2}}\big)$, with $K_1 =  2 c^3 -9 (bcd - 3 b^2 e - 3 d^2 + 8c e)$, and finally,
\begin{equation}
\begin{split}
  b &= - \sum_i \Delta m_{i1}^2 - A_\text{CC} - A_\text{NC} \,, \\[2mm]
  c &= A_\text{CC}A_\text{NC} 
  +  \sum_i \Delta m_{i1}^2 \left[A_\text{CC}\left(1- |U_{ei}|^2\right) + A_\text{NC} \left(1- |U_{si}|^2\right)\right] 
  +  \sum_{i<j} \Delta m_{i1}^2\Delta m_{j1}^2 \,,\\[2mm]
  d &= A_\text{CC}A_\text{NC}\sum_i \Delta m_{i1}^2 \left(|U_{ei}|^2 + |U_{si}|^2 - 1\right) 
  - \Delta m_{21}^2 \Delta m_{31}^2 \Delta m_{41}^2 \\
  &\quad
  -  \sum_{i<j} \sum_k (1-\delta_{ik})(1-\delta_{jk})\Delta m_{i1}^2\Delta m_{j1}^2 
    \left(A_\text{CC} |U_{ek}|^2 + A_\text{NC} |U_{sk}|^2\right)
  \,,\\[2mm]
  e &= A_\text{CC}A_\text{NC} \sum_{i<j}\sum_{k<l}
  (1-\delta_{ik})(1-\delta_{jk})(1-\delta_{il})(1-\delta_{jl})\Delta m_{i1}^2\Delta m_{j1}^2
  |U_{ek}U_{sl}-U_{el}U_{sk}|^2\\
  &\quad
  +\Delta m_{21}^2 \Delta m_{31}^2 \Delta m_{41}^2 \left(A_\text{CC} |U_{e1}|^2 + A_\text{NC} |U_{s1}|^2\right) \,.
\end{split}
\raisetag{1.1\normalbaselineskip}
\end{equation}

The mass-squared differences in matter $\Delta \tilde m^2_{i1} = \hat \Delta m^2_{i1} - \hat \Delta m^2_{11}$ follow. For a neutrino spectrum with normal ordering,
\begin{equation}
\begin{split}
  \Delta \tilde m^2_{21} &= \sqrt{-4S^2-2p+\frac{q}{S}}\,,\\[1mm]
  \Delta \tilde m^2_{31} &= 2S + \frac{1}{2} \left( \sqrt{-4S^2-2p+\frac{q}{S}} -  \sqrt{-4S^2-2p-\frac{q}{S}}\right)\,,\\[1mm]
  \Delta \tilde m^2_{41} &= 2S + \frac{1}{2} \left( \sqrt{-4S^2-2p+\frac{q}{S}} +  \sqrt{-4S^2-2p-\frac{q}{S}}\right)\,.
\end{split}
\label{eq:mtNO}
\end{equation}
In the case of inverted ordering, one has instead
\begin{equation}
\begin{split}
  \Delta \tilde m^2_{21} &= 2S - \frac{1}{2} \left( \sqrt{-4S^2-2p+\frac{q}{S}} + \sqrt{-4S^2-2p-\frac{q}{S}}\right)\,,\\[1mm]
  \Delta \tilde m^2_{31} &= -\sqrt{-4S^2-2p+\frac{q}{S}}\,,\\[1mm]
  \Delta \tilde m^2_{41} &= 2S - \frac{1}{2} \left( \sqrt{-4S^2-2p+\frac{q}{S}} - \sqrt{-4S^2-2p-\frac{q}{S}}\right)\,.
\end{split}
\label{eq:mtIO}
\end{equation}

\subsection{Mixing in matter}

It is possible to obtain analytic expressions for the products $\tilde U_{\alpha i} \tilde U_{\beta i}^*$ ($i=1,...,4$; $\alpha,\beta=e,\mu,\tau,s$) of mixing matrix elements in matter~\cite{Zhang:2006yq,Li:2018ezt}. Following ref.~\cite{Li:2018ezt}, we take as a starting point the relation
\begin{equation}
    \tilde U\, \tilde d\, \tilde U^\dagger
    \,=\, \underbrace{ U\,d\,U^\dagger+ \mathcal{A}}_{2 E H^\text{mat}} - \hat \Delta m^2_{11} \id
    \,\equiv\, \mathcal{M}\,.
    \label{eq:M}
\end{equation}
where we have defined
$\tilde d\equiv \diag(0,\Delta\tilde m^2_{21},\Delta\tilde m^2_{31},\Delta\tilde m^2_{41})$, 
$d\equiv \diag(0,\Delta m^2_{21}, \Delta m^2_{31}, \Delta m^2_{41})$,
and $\mathcal{A} \equiv \diag(A_\text{CC},0,$ $0,A_\text{NC})$. Taking \cref{eq:M}, its square and its cube, one finds the matrix equation
\begin{equation}
\begin{pmatrix}
\Delta \tilde m^2_{21} &  \Delta \tilde m^2_{31} &  \Delta \tilde m^2_{41} \\[1mm]
\left( \Delta \tilde m^2_{21}\right)^2 & \left( \Delta \tilde m^2_{31}\right)^2 & \left( \Delta \tilde m^2_{41}\right)^2 \\[1mm]
\left( \Delta \tilde m^2_{21}\right)^3 & \left( \Delta \tilde m^2_{31}\right)^3 & \left( \Delta \tilde m^2_{41}\right)^3
\end{pmatrix} 
\begin{pmatrix}
\tilde U_{\alpha 2}\tilde U^*_{\beta 2} \\[1mm]
\tilde U_{\alpha 3}\tilde U^*_{\beta 3} \\[1mm]
\tilde U_{\alpha 4}\tilde U^*_{\beta 4} 
\end{pmatrix} \,=\,
\begin{pmatrix}
\mathcal{M}_{\alpha\beta} \\[1mm]
\mathcal{M}_{\alpha\gamma}\mathcal{M}_{\gamma\beta} \\[1mm]
\mathcal{M}_{\alpha\gamma}\mathcal{M}_{\gamma\delta}\mathcal{M}_{\delta\beta}
\end{pmatrix} \,,
\end{equation}
whose solution is
\begin{equation}
\begin{split}
\tilde U_{\alpha 2}\tilde U^*_{\beta 2}\,&=\,
\frac{\Delta \tilde m^2_{31} \Delta \tilde m^2_{41} \mathcal{M}_{\alpha\beta}-\left(\Delta \tilde m^2_{31} +\Delta \tilde m^2_{41}\right) \mathcal{M}_{\alpha\gamma}\mathcal{M}_{\gamma\beta}+\mathcal{M}_{\alpha\gamma}\mathcal{M}_{\gamma\delta}\mathcal{M}_{\delta\beta}}{\Delta \tilde m^2_{21} \left(\Delta \tilde m^2_{21}-\Delta \tilde m^2_{31}\right) \left(\Delta \tilde m^2_{21}-\Delta \tilde m^2_{41}\right)}
\,, \\[2mm]
\tilde U_{\alpha 3}\tilde U^*_{\beta 3}\,&=\,
\frac{\Delta \tilde m^2_{21} \Delta \tilde m^2_{41} \mathcal{M}_{\alpha\beta}-\left(\Delta \tilde m^2_{21} + \Delta \tilde m^2_{41}\right) \mathcal{M}_{\alpha\gamma}\mathcal{M}_{\gamma\beta}+\mathcal{M}_{\alpha\gamma}\mathcal{M}_{\gamma\delta}\mathcal{M}_{\delta\beta}}{\Delta \tilde m^2_{31} (\Delta \tilde m^2_{31}-\Delta \tilde m^2_{21}) (\Delta \tilde m^2_{31}-\Delta \tilde m^2_{41})}\,, \\[2mm]
\tilde U_{\alpha 4}\tilde U^*_{\beta 4}\,&=\,
\frac{\Delta \tilde m^2_{21} \Delta \tilde m^2_{31} \mathcal{M}_{\alpha\beta} - \left(\Delta \tilde m^2_{21} + \Delta \tilde m^2_{31}\right) \mathcal{M}_{\alpha\gamma}\mathcal{M}_{\gamma\beta}+\mathcal{M}_{\alpha\gamma}\mathcal{M}_{\gamma\delta}\mathcal{M}_{\delta\beta}}{\Delta \tilde m^2_{41} (\Delta \tilde m^2_{41}-\Delta \tilde m^2_{21}) (\Delta \tilde m^2_{41}-\Delta \tilde m^2_{31})}\,.
\end{split}
\end{equation}
After quite a bit of algebra, and taking into account the unitarity relation 
$\delta_{\alpha\beta} = \tilde U_{\alpha 1}\tilde U^*_{\beta 1}
+\tilde U_{\alpha 2}\tilde U^*_{\beta 2}
+\tilde U_{\alpha 3}\tilde U^*_{\beta 3}
+\tilde U_{\alpha 4}\tilde U^*_{\beta 4} $, one reaches the main result
\begin{equation}
    \tilde U_{\alpha i}\tilde U^*_{\beta i}\,=\, \frac{1}{\prod_{k\neq i} \Delta \tilde m^2_{ik}}
\left[\sum_j F_{\alpha\beta}^{ij} U_{\alpha j}U^*_{\beta j} + C_{\alpha\beta}\right]\,,
\label{eq:UtUts}
\end{equation}
with
\begin{equation}
\begin{split}
  F_{\alpha\beta}^{ij} \,&=\,
  \prod_{k\neq i} \left(\frac{\aa+\bb}{2} - \mh{kj}\right)
 -\left(1-\delta_{\alpha\beta}\right)\prod_{k\neq i} \left(\frac{\aa+\bb}{2} - \mh{k1}\right)
 \\
 & \quad
 +\m{j1}\left(\frac{\aa-\bb}{2}\right)^2\,, 
\\
C_{\alpha\beta} \,&=\, 
- \frac{1}{2}\sum_{m,n,\gamma}\left(\m{mn}\right)^2 \A{\gamma \gamma}
U_{\alpha m}U_{\gamma m}^* U_{\gamma n}U_{\beta n}^* \,.
\end{split}
\label{eq:FC}
\end{equation}
Recall that $\mathcal{A} = \diag(A_\text{CC},0,$ $0,A_\text{NC})$ and $\hat \Delta m_{i1}^2 = \Delta \tilde{m}_{i1}^2 + \hat \Delta m_{11}^2$, while the $\Delta \tilde m^2_{ij}$ can be obtained from \mbox{\cref{eq:mtNO,eq:mtIO}}.
One has further defined $\hat \Delta m_{ij}^2 \equiv \hat \Delta m_{i1}^2 - \Delta m_{j1}^2$.

\vskip 2mm
We have checked that these results are valid for all $i=1,...,4$ and $\alpha,\beta=e,\mu,\tau,s$, comprising both the cases $\alpha = \beta$ and $\alpha \neq \beta$.
Our results are compatible with those of ref.~\cite{Li:2018ezt},%
\footnote{
There is a typo $U_{sk}U_{sl}^* \to U_{sk}^*U_{sl}$ in eq.~(C.10) and consequently~(3.10) of~\cite{Li:2018ezt}. Also, in eq.~(C.9) therein one should be careful to interpret the potentially ambiguous sum $\sum_{k,l;k\neq l \neq i}$ as $\sum_{k>l; k\neq i; l \neq i}$.
}
which correct those of~\cite{Zhang:2006yq}.
Note that while our definitions of $F_{\alpha\beta}^{ij}$ and $C_{\alpha\beta}$ do not directly match those in~\cite{Li:2018ezt}, due to a different way of grouping the terms, the complete expression~\eqref{eq:UtUts} does after it is expanded.

\section{Approximate short-baseline probabilities}
\label{app:approx}

In this appendix, we explicitly derive the approximate oscillation probability formula of \cref{eq:SBL}, used for the estimation of ND event rates. 
We start by neglecting matter effects in short-baseline oscillations. Then, for a single baseline $L$ in vacuum, which matches \cref{eq:PfilterFD} after dropping tildes, one has
\begin{equation}
\langle P_{\alpha\beta}^\text{SBL}(L,E)\rangle 
\,\simeq\,
\sum_{j,j'}
U_{\beta j}
U^\dagger_{j\alpha}
U_{\alpha j'}
U^\dagger_{j'\beta}
\exp\left(-i\frac{\Delta m^2_{j j'}L}{2E}\right) \, \text{l.p.f.}_{j j'}\,,
\end{equation}
with the replacement $U\rightarrow U^{\ast}$ for antineutrinos.
Here, l.p.f.~denotes the low-pass filter factor,
\begin{equation}
\text{l.p.f.}_{j j'} \equiv \exp\left[-\frac{\sigma^2_E}{2E^2}\left(\frac{\Delta m^2_{j j'}L}{2E}\right)^2\right].    
\end{equation}
Neglecting the solar and atmospheric mass-squared differences with respect to the sterile one and taking $\Delta m^2_{41}\simeq \Delta m^2_{42}\simeq \Delta m^2_{43}$, the above probabilities can be further approximated as
\begin{equation}
\begin{aligned}
\langle P_{\alpha\beta}^\text{SBL}\rangle
&\simeq \sum_{j,j'=1}^3 U_{\beta j}U^{\dagger}_{j\alpha}U_{\alpha j'}U^{\dagger}_{j'\beta}  \\
&\quad + U_{\beta 4}U^{\dagger}_{4\alpha}
\left[\,\sum_{j'=1}^3 U_{\alpha j'}U^{\dagger}_{j'\beta} \exp\left(-i\frac{\Delta m^2_{41}L}{2E}\right)\text{l.p.f.}_{41}
+U_{\alpha 4}U^{\dagger}_{4\beta}\,\right] \\
&\quad+ U_{\alpha 4}U^{\dagger}_{4\beta}\sum_{j=1}^3 U_{\beta j}U^{\dagger}_{j\alpha}\exp\left(i\frac{\Delta m^2_{41}L}{2E}\right)\text{l.p.f.}_{41}\,.
\label{eq:vac1}
\end{aligned}
\end{equation}
Using the unitarity condition $\sum_{j=1}^4 U_{\alpha j}U^{\dagger}_{j\beta}=\delta_{\alpha\beta}$,
the first line in \mbox{\cref{eq:vac1}} becomes
\begin{equation}
\delta_{\alpha\beta}-\delta_{\alpha\beta}\left(U_{\alpha 4}U^\dagger_{4\beta}+U_{\beta 4}U^\dagger_{4\alpha}\right)+|U_{\alpha 4}|^2|U_{\beta 4}|^2\,, 
\label{eq:vac2}
\end{equation}
while the second and third lines give
\begin{equation}
\begin{aligned}
&\left\{
\delta_{\alpha\beta} \left[
 U^\dagger_{4\alpha}U_{\beta 4} \exp\left(-i\frac{\Delta m^2_{41}L}{2E}\right)
+U_{\alpha 4}U^\dagger_{4\beta} \exp\left(i\frac{\Delta m^2_{41}L}{2E}\right)
\right] 
\right. \\ & \left.
-2|U_{\alpha 4}|^2|U_{\beta 4}|^2 \cos\left(\frac{\Delta m^2_{41}L}{2E}\right)
\right\}\,\text{l.p.f.}_{41} \,+\, |U_{\alpha 4}|^2|U_{\beta 4}|^2\,.
\end{aligned}
\label{eq:vac3}
\end{equation}
Both \eqref{eq:vac2} and the first line of \eqref{eq:vac3} can be further simplified using the properties of $\delta_{\alpha\beta}$.
Then, by adding these terms, \cref{eq:SBL} follows,
\begin{equation*}
\begin{aligned}
\langle P_{\alpha\beta}^\text{SBL}\rangle
&\simeq
\delta_{\alpha\beta}
- 2\delta_{\alpha\beta}|U_{\alpha 4}|^2 
+2|U_{\alpha 4}|^2|U_{\beta 4}|^2\left[1-\cos\left(\frac{\Delta m^2_{41} L}{2E}\right)\text{l.p.f.}_{41}\right] 
\\ &\quad
+ 2\delta_{\alpha\beta}|U_{\alpha 4}|^2 \cos\left(\frac{\Delta m^2_{41}L}{2E}\right)\text{l.p.f.}_{41} \\[2mm]
&=
\delta_{\alpha\beta}-2|U_{\alpha 4}|^2 \left(\delta_{\alpha\beta}-|U_{\beta 4}|^2\right)
\left\{1-\cos\left(\frac{\Delta m^2_{41}L}{2E}\right)\exp\left[-\frac{\sigma^2_E}{2E^2}\left(\frac{\Delta m^2_{41}L}{2E}\right)^2\right]\right\}\,.
\end{aligned}
\end{equation*}
In the limit $\sigma_E = 0$, one recovers the usual SBL result (cf.~ref.~\cite{Giunti:2019aiy}).

\def\bibsection{\section*{\refname}}

\bibliographystyle{JHEPwithnote}
\bibliography{bibliography}

\end{document}